\documentclass[conference,compsoc]{IEEEtran}
\usepackage{graphicx} %
\usepackage{amsmath, amssymb, amsthm}
\usepackage{xcolor}
\usepackage{hyperref}
\usepackage{booktabs}
\usepackage{subcaption}
\usepackage[capitalize]{cleveref}
\usepackage[numbers]{natbib}
\usepackage{xurl}
\usepackage{xspace}
\usepackage{caption}
\usepackage{svg}
\usepackage{enumitem}
\usepackage{multirow}
\usepackage{array}

\newcommand{\modfilter}{\mathcal{C}}
\newcommand{\surrmodel}{\mathcal{F}}
\newcommand{\methodname}{{\sc Mirage}\xspace}

\crefname{equation}{Eq.}{Eqs.}

\title{\textsc{MIRAGE}: Protecting against Malicious Image Editing via False Moderation}

\author{
    \IEEEauthorblockN{
        Anshul Nasery\textsuperscript{1},
        Ramnath Kumar\textsuperscript{2},
        Cho-Jui Hsieh\textsuperscript{2,3},
        Sewoong Oh\textsuperscript{1}
    }
    \IEEEauthorblockA{
        \textsuperscript{1}University of Washington
        \hspace{1.5em}
        \textsuperscript{2}University of California, Los Angeles
        \hspace{1.5em}
        \textsuperscript{3}Arena
        \\[0.3em]
        \{anasery, sewoong\}@cs.washington.edu,
        \{ramnathk, chohsieh\}@cs.ucla.edu
    }
}

\begin{document}

\maketitle

\begin{abstract}
The proliferation of AI-powered image editing systems raises serious concerns because it allows personal images to be arbitrarily manipulated at scale, with minimal effort, and a lower barrier to entry.
Prior work on \textit{image immunization} adds imperceptible perturbations to an image to protect against unauthorized manipulations. However, these methods usually require access to the model weights and the image manipulating prompt. This significantly limits their use, especially against powerful commercial image-editors such as GPT-Image, Gemini Flash Image (Nano Banana), and Grok Imagine.
To address this, we take a system-level view of the problem and identify a previously unexplored attack surface common to all major commercial image editing systems: pre-generation safety moderation. 
Rather than disrupting the generative model itself, %
we propose to immunize images by causing these moderation classifiers to flag images as policy-violating, triggering an automatic refusal regardless of the editing prompt. We operationalize this by adding adversarial perturbations to align our image to policy-violating concepts in the representation space of an ensemble of open-source embedding and moderation models. We call our method \methodname, which stands for Moderation Induced Resistance Against Generative Editing.
We evaluate \methodname against multiple closed-source image editing APIs and demonstrate success rates of more than 88\%. 
Our approach is simple, prompt-agnostic, and effective, offering a practical path towards protecting personal images from unauthorized AI-powered editing.
\end{abstract}

\section{Introduction}

The rapid advancement of AI-powered image editing has fundamentally lowered the barrier to realistic, large-scale photo-manipulation. Systems such as OpenAI's GPT-Image \cite{gpt-image-2}, Google's Gemini \cite{gemini-3.1-flash-image}, and xAI's Grok \cite{grok-imagine-image} allow any user to upload any image and apply sophisticated, semantically meaningful edits through natural language prompts.  This includes  changing a person's appearance, altering their surroundings, or placing them in entirely fabricated scenarios. 
While these capabilities can serve genuine creative tasks, they simultaneously enable a new class of consent violations;  individuals' personal photographs and images can be maliciously edited without their knowledge or agreement, 
enabling fraud \cite{diresta2024spammers}, harassment \cite{gentleman_grok_xai_lawsuit_2026,digitalhate,digitalhate2,digitalhate3,digitalhate4}, defamation \cite{defamation,defamation1},  political disinformation \cite{political,political1},  and identity manipulation \cite{impersonation}, at a scale previously impossible.

In response to this escalating threat, the research community has proposed a series  of 
solutions for {\em image immunization}: adding  visually imperceptible perturbations to an image that cause the image-editing models to fail. %
However, all existing methods suffer from the lack of {\bf transferability} in both the models and the prompts. Popular  
methods, including  PhotoGuard~\cite{salman2023photoguard}, EditShield~\cite{chen2024editshield}, and others~\cite{guo2024posterior,dong2025semantic,lo2024distraction,ozden2024diffvax}, require white-box access to  the model weights and also the prompt, that are to be used by the perpetrator. 
Image perturbations optimized for one open-source model and for one prompt do not reliably transfer to different models and prompts. 
In Figure~\ref{fig:main_results}, for example, we empirically demonstrate   that none of the existing immunization methods are effective against closed-source black-box commercial APIs. This leaves the independent individuals who own and share  images and photos  almost entirely unprotected against an adversary with easy access to  the most powerful and widely-used  systems from companies like OpenAI, Google, and xAI. 

This leads to a fundamental question in making image immunization practical: {\em is there an image immunization method that can protect against the frontier AI-powered image editors like GPT-Image \cite{gpt-image-2}, Google's Nano Banana \cite{gemini-3.1-flash-image}, and Grok Imagine \cite{grok-imagine-image}?} Further, we want a single immunized image to universally work for all these powerful systems and for all image editing prompts, simultaneously.

One could attempt to use black-box adversarial attacks, which rely on zeroth-order optimization methods, to directly immunize images against editing by frontier API systems, e.g., \cite{andriushchenko2020squareattackqueryefficientblackbox, cheng2018queryefficienthardlabelblackboxattackan,ilyas2018black,chen2017zoo,brendel2017decision,tu2019autozoom}. However, such methods have large, dimension-dependent query complexity \cite{duchi2015optimal}. For images, this means that one needs to query the API thousands of times. %
This is prohibitively expensive, both in terms of cost (over 50 USD per image) and time (about 15-30 seconds per API call). Alternatively, universal perturbation approaches designed to transfer to any image-editors, such as~\citep{ahn2025nearly,kim2025blurguard}, have been introduced. However, these techniques require a large ensemble of surrogate models that includes models with similar architecture to the target models. With commercial, black-box image editing systems, one cannot make any claims about the architecture of the underlying model, hence such universal perturbations are ineffective in immunizing images, as we see in \cref{fig:main_results}. We also provide qualitative examples in Figs.~\ref{fig:baseline-photoguard-qualitative-results} and \ref{fig:baseline-tdae-qualitative-results}.
\begin{figure*}
\includegraphics[width=1.0\linewidth]{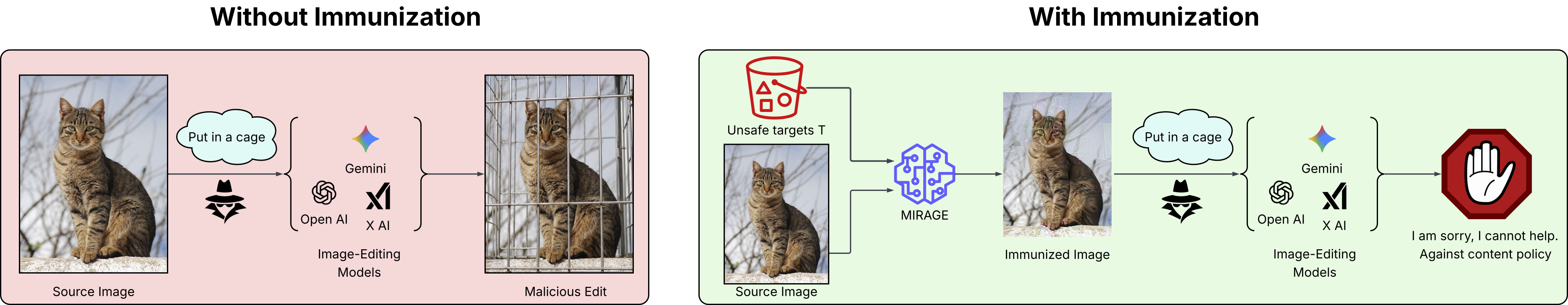}
    \caption{\textbf{Overview of the image immunization framework}.
    \textit{(Left - Without Immunization):} Given a source image (a cat sitting outdoors), an adversary can pair it with a (potentially malicious) instruction (\textit{``Put in a cage''}) and feed it to a black-box image-editing model (e.g., Gemini, OpenAI, or xAI). The model complies with the instruction, producing an image (in which the cat appears behind cage bars).
    \textit{(Right - With Immunization):} To protect against such edits, an image owner could \textit{immunize} their image. Our immunization pipeline takes in a set of unsafe target images $\mathcal{T}$ (representing moderation triggering inputs) along with the source image. It perturbs the source image to produce a visually similar immunized image. When this immunized image is subsequently submitted to the same image-editing model alongside the same  instruction, the model refuses to comply and returns a content policy violation message (e.g. \textit{``I am sorry, I cannot help.''}).}
    \label{fig:hero_fig}
\end{figure*}

To this end, we take a system-level view of current frontier closed-source  image-editing systems and identify a fundamental structural property: 
all major commercial APIs implement  responsible-use policies enforced by a 
safety classifier. Before any edits are produced, each system passes the input image through a moderation pipeline; if the image is flagged as policy-violating, the system refuses to generate output entirely,  regardless of the prompt. We propose to turn this safety mechanism, originally intended to prevent misuse, into a defensive tool for the image owners. 
We call this new paradigm  \textit{false moderation}: adding visually imperceptible  perturbations to an image that cause the safety classifier of a commercial image-editing API to flag the immunized image as policy-violating, producing a refusal that is prompt-agnostic.

Attacking closed-source safety classifiers poses a challenge: we do not have direct access to them. We address this through transfer-based adversarial attacks over an ensemble of open-source models~\cite{zhang2026launderingaiauthorityadversarial}, including CLIP-style vision-language encoders, and open moderation models such as ShieldGemma\cite{zeng2025shieldgemma2robusttractable}. The main idea of our method is summarized in \cref{fig:hero_fig}. We first investigate the terms of use of the systems we want to immunize against, xAI's Grok~\cite{xai_acceptable_use_policy_2025}, Google's Gemini~\cite{google_gemini_api_terms_2026}, OpenAI's GPT-Image~\cite{openai_usage_policies_2025}, and extract text descriptions or images associated with policy-violating content categories. We use these to guide the image perturbation to make the image seem more unsafe. The resulting perturbation is small in magnitude, visually imperceptible, and causes the target API to refuse to produce any edit. We term our approach \methodname, which stands for \textbf{M}oderation \textbf{I}nduced \textbf{R}esistance \textbf{A}gainst \textbf{G}enerative \textbf{E}diting.

In summary, we make the following contributions:

\begin{itemize}
    \item We identify the pre-generation safety classifiers as a novel and practically effective attack surface for image immunization against closed-source AI-powered image editing systems. We propose \methodname, a method that generates adversarial perturbations causing commercial APIs to refuse to edit an image (\cref{sec:method}).
    \item %
    We empirically test \methodname  
    against three commercial frontier image-editing APIs: OpenAI GPT-Image~\cite{gpt-image-2}, Google Gemini~\cite{gemini-3.1-flash-image}, and xAI Grok~\cite{grok-imagine-image}. 
    We show that \methodname achieves immunization success rates of more than 88\% on all systems, while all prior immunization baselines achieve very low success rates (\cref{fig:main_results}). 
    \item We provide an ablation study on each component of our design, which demonstrates that the success of the proposed immunization critically relies on ($i$) using an ensemble of open-source embedding models, ($ii$) combining both global and local views of the input image, and ($iii$) the choice of target concept selected from the acceptable use policy in the terms of use
    (Section \ref{sec:results-ablations}).
    \item We  evaluate robustness against diverse  adversaries from weak to strong (defined in \Cref{sec:threat}).  
    We demonstrate our immunization remains largely effective against a weak adversary who applies standard image processing operations to bypass immunization (\cref{fig:robustness}). Robustness results against stronger adversaries suggest that while it is an open problem how to completely protect against strong adversaries, \methodname is the most effective method in raising the cost of the adversary who attempts to apply harmful manipulations (\cref{fig:strong-robustness}).
\end{itemize}

\section{Preliminaries}

We define two principals: the \textit{adversary}, who seeks to manipulate images posted online without the consent of their respective image owners, and the \textit{image owner}, who wishes to share images without the risk of malicious manipulation.

\subsection{Threat model: unauthorized  manipulation}
\label{sec:threat}

The adversary has access to an image $x\in{\cal X}$ and a proprietary black-box AI-powered image-editor $f:{\cal X}\to{\cal X}$, such as GPT-image-2 \cite{gpt-image-2}, Gemini-3.1-flash-image \cite{gemini-3.1-flash-image}, or Grok-imagine-image \cite{grok-imagine-image}. The adversary's goal is to transform $x$ into a target image $y=f(x)$ without the consent of the image owner. This captures {\em unauthorized image manipulation} that can violate copyright or generate harmful or misleading images. Note that typical image-editors take in additional inputs such as text prompts or style reference images, which we intentionally omit in the definition of $f$ assuming it is clear from the context. This abstraction also signifies our assumption that the adversary, in response to immunization techniques, ($i$) can preprocess the shared image $x$ before feeding it into the image editor $f$, but ($ii$) does not change the auxiliary text or image inputs attempting to jailbreak the immunization. 

We consider two types of preprocessing capabilities of the adversary: weak and strong. A weak adversary can only use classical image processing transformations to aim to bypass the immunization. Under the stronger threat model, the adversary also has white-box access to some open-source models for image embedding, editing, or generation. They can hence choose to edit or purify~\cite{nie2022DiffPure} the image locally. We measure the robustness of our technique under both types of adversaries in \Cref{sec:results}. 

We focus our main results on proprietary APIs since ($i$) existing image immunization techniques such as Photoguard \cite{salman2023photoguard} already work well against open-source image-editors,
and ($ii$) proprietary image-editors provide significantly higher utility to the adversary; proprietary image-editors produce significantly higher quality outputs. We empirically show that existing white-box immunization methods like Photoguard do not transfer to frontier AI-powered image-editors in Figure~\ref{fig:main_results}. To the best of our knowledge, we are the first to successfully immunize images against black-box, proprietary image-editors.

\subsection{Problem formulation: image immunization} 
\label{sec:problem} 

The image owner wants to post online  the original image $x$, knows that there is an adversary attempting to manipulate the image in a harmful way, but does not know which proprietary image-editor the adversary has access to. Further, neither the adversary nor the image owner knows the inner workings of each proprietary model other than what is publicly released. For example, in our design, we only assume access to the terms-of-use  of multiple proprietary AI-powered image-editing APIs, including xAI's Grok~\cite{xai_acceptable_use_policy_2025}, Google's Gemini~\cite{google_gemini_api_terms_2026}, OpenAI's GPT-Image~\cite{openai_usage_policies_2025}. In  \methodname, we use these terms-of-use documents to figure out what concept constitutes safety violation, thus triggering a moderator to refuse to produce the image output.
We further assume that the image owner has access to any open-source models including, in particular, image-text embedding models such as CLIP \cite{radford2021learningtransferablevisualmodels}, OpenCLIP \cite{Cherti_2023}, and DataComp \cite{gadre2023datacompsearchgenerationmultimodal}. Our method uses these open-source image embedding models to find similarities between  perturbations in the image space and unsafe concepts in the latent space.

The image owner's goal is to raise the cost of malicious AI-powered image editing by posting a perturbed version $\hat{x}$ instead of $x$ such that ($i$) it is visually close to the original image as measured by some image distance metric $d(x,\hat{x})$, and ($ii$) it renders image transformation ineffective, e.g.,  $d(f(\hat{x}), f(x))$ is large. An adversary who knows how such {\em image immunization scheme} works can attempt to bypass it by preprocessing the shared image, as in the weak and strong adversaries mentioned above. We test the robustness of our immunization against both weaker/compound adversaries, and stronger adversaries in Figure~\ref{fig:robustness} and Figure~\ref{fig:strong-robustness} respectively. The goal in such a scenario is to force the adversary to spend additional resources, making them less incentivized to maliciously edit images without prior consent. 

To achieve the above goals, we introduce \textbf{M}oderation \textbf{I}nduced \textbf{R}esistance \textbf{A}gainst \textbf{G}enerative \textbf{E}diting (\methodname). Rather than attacking the generative model $f$ directly, which is inaccessible in the black-box setting, the key innovation in \methodname~is to exploit a structural property shared across all major commercial image-editing APIs: a safety moderator $C$ that refuses to produce output when the input image is flagged as policy-violating for a myriad of reasons. By adding an imperceptible adversarial perturbation that causes $C(\hat{x})$ to exceed the refusal threshold, \methodname~renders $\hat{x}$ immune to editing regardless of the adversary's prompt (see \Cref{fig:hero_fig}). We describe our method in full in Section~\ref{sec:method}. 

\section{Related work}
\label{sec:related}

We situate our work at the intersection of three lines of research: ($i$) image immunization against AI-powered editing, ($ii$)~adversarial attacks repurposed for defensive goals, and ($iii$)~black-box attacks on proprietary models. While each contributes partial foundations, none addresses the problem of immunizing images against black-box production-grade image-editors---the gap our method is designed to fill.

\subsection{Image immunization}
As image editing models become more powerful, the potential of their misuse has also increased. As a countermeasure, a recent line of works has sought to disrupt or raise the cost of malicious AI-powered image editing by harnessing adversarial examples~\cite{goodfellow2015explainingharnessingadversarialexamples} for diffusion models. Photoguard~\cite{salman2023photoguard} proposes adding a small perturbation $\delta$ to the  original image $x$  such that when the perturbed image is passed to the diffusion-based editing model $f$, the output is visually degraded. To compute the perturbation, Photoguard performs PGD~\cite{madry2017towards} to minimize $||f(x+\delta) - y_{\text{tgt}}||^2_2$, where $y_{\text{tgt}}$ is an arbitrary target image chosen by the image owner (e.g. a completely black image). Crucially, this requires computing $\nabla_\delta f(x+\delta)$, limiting its application to white-box image-editors.  

One downside of Photoguard is the cost of backpropagating through the entire diffusion model $f$. To mitigate the cost, prior works~\cite{salman2023photoguard, chen2024editshield, guo2024posterior} have focused on the popular latent diffusion based image editing models~\cite{rombach2022highresolutionimagesynthesislatent} of the form $f(x) = g(\mathcal{E}(x))$, where $\mathcal{E}(\cdot)$ is an encoder of a VAE and $g(\cdot)$ is the diffusion model in the lower-dimensional latent space. 
Concretely, an alternate version of Photoguard also from  \cite{salman2023photoguard}  tries to maximize the distance between $x+\delta$ and $x$ in the latent space of the VAE, i.e., $||\mathcal{E}(x+\delta) - \mathcal{E}(x)||_2$,  used for the latent diffusion model of interest. Follow-up works have modified this objective to collapse the posterior of the VAE~\cite{guo2024posterior}. A related line of works aim to create a mismatch between attention matrices~\cite{zhang2025dual,lo2024distraction} or intermediate features~\cite{dong2025semantic} of $x+\delta$ and $x$ in the diffusion transformer used by $f$.  

Another source of wasted resources is that the perturbation $\delta$ needs to be reinvented for each image $x$.  More recent works address this by borrowing techniques from the adversarial attacks literature, such as using augmentations~\cite{chen2024editshield}, masking~\cite{choi2025diffusionguard}, universal adversarial attacks to get a single perturbation $\delta$ that works across multiple images~\cite{zhang2026towards, kim2025blurguard}, or using a learnt network~\cite{ahn2025nearly} to generate $\hat{x}$ from $x$.
However, note that all these methods require complete white-box access to the model that the image is to be immunized against.
Further, unlike adversarial examples against classification models, we note that these perturbations do not transfer from one diffusion model to another. Recent works have attempted to regularize the perturbation to encourage such transfer~\cite{zhang2026towards}, however, as we show in \Cref{sec:results}, such perturbations are ineffective against production black-box image editing systems. This is especially concerning since most powerful image-editing models are black-box, i.e., the user only has API access to these. As a result, existing image immunization methods cannot protect against such models, as we show in Figure~\ref{fig:main_results}.

\subsection{Adversarial attacks for good}

The field of image immunization~\cite{salman2023photoguard} uses adversarial attacks to prevent malicious edits to user images. There have been other lines of work reframing techniques from adversarial machine learning for defending against unauthorized AI usages. Nightshade~\cite{shan2024nightshade} and Glaze~\cite{shan2023glaze} leverage data poisoning attacks to protect style mimicry of artists' work by generative AI models. Model fingerprinting~\cite{zhang2018protecting,xu2024instructional, nasery2026scalable,nasery2025robust} uses backdoor attacks to detect misuse of open-weights models. Other works~\cite{shao2026leave} also construct adversarial examples on white-box visual question answering (VQA) models to prevent them from answering questions on a user's private images. 
Although image immunization is posed as an adversarial attack problem, our key technique is novel:  turning  safety moderation of proprietary models against those proprietary systems to make image sharing safe. 

\subsection{Attacking proprietary models} %
Popular approaches to attack proprietary machine learning APIs rely on either black-box optimization techniques~\cite{andriushchenko2020squareattackqueryefficientblackbox, cheng2018queryefficienthardlabelblackboxattackan} or surrogate-based transfer~\cite{liu2016delving,lord2022attacking,zhang2025maameticulousadversarialattack,jia2025adversarialattacksclosedsourcemllms}. However, these approaches face a fundamental scalability challenge when applied to image immunization. Black-box gradient estimation methods such as NES~\cite{ilyas2018black}, and ZOO~\cite{chen2017zoo} typically require thousands to tens of thousands of queries even for targeted attacks in the relatively simpler classification setting~\cite{andriushchenko2020squareattackqueryefficientblackbox}, which involves estimating the gradient $\nabla_\delta \mathcal{L}$ of a scalar loss signal (a logit or hard label).

Commercial image-editing endpoints such as those built around GPT-Image and Google Gemini incur latencies of 15-30 seconds per forward pass, meaning even modest budget of 1{,}000 queries translates to roughly 6-8 hours of wall-clock time per image, clearly impractical for end users seeking real-time protection. Prior work on hard-label black-box attacks~\cite{cheng2018queryefficienthardlabelblackboxattackan, brendel2017decision} has made progress on query efficiency, but still remains intractable for frontier image generation models; it is evaluated only in classification settings with fast, cheap oracles.

For large vision-language models (VLMs), prior work has further shown that finding transferable adversarial examples is challenging~\cite{schaeffer2025failures}, prompting more sophisticated ensemble-based approaches with intermediate feature alignment~\cite{zhang2025maameticulousadversarialattack,jia2025adversarialattacksclosedsourcemllms}. The closest approach to ours is a concurrent work~\cite{zhang2026launderingaiauthorityadversarial}  which uses a CLIP ensemble to construct perturbations which can \textit{bypass the moderation} and decision making pathways of commercial VLMs. This enables the production of output that is policy-violating, e.g. producing fake news about real people. Our work on the other hand perturbs input images which induce content-moderation filters to \textit{refuse} outputs on otherwise ``benign" images.   

\section{Methodology: \methodname}
\label{sec:method}

\begin{figure*}
    \includegraphics[width=0.95\linewidth]{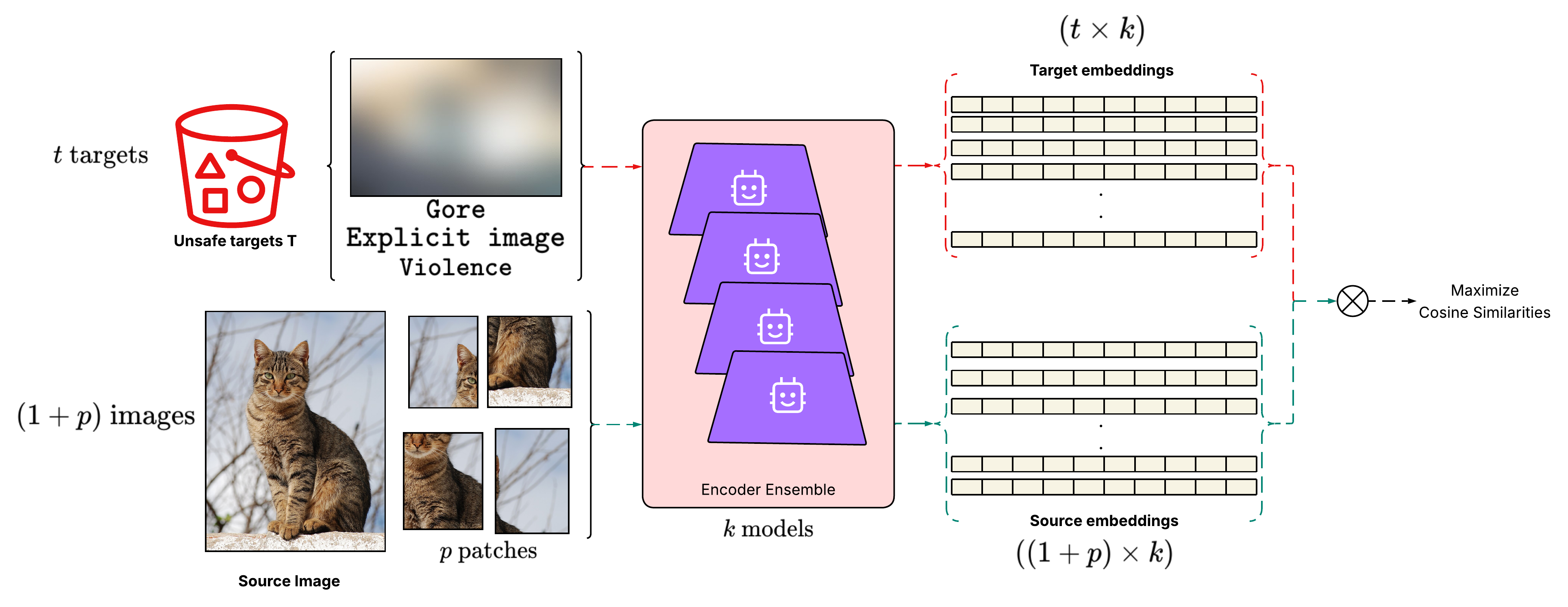}
    \caption{\textbf{\methodname : Immunization objective via embedding similarity.}
The immunization pipeline computes adversarial perturbations to the source
image by maximizing its alignment with a set of unsafe target embeddings. The source image and each image/text in the unsafe target set
$\mathcal{T}$ (e.g., gore, violence, sexually explicit content, etc.) are independently encoded by many frozen image/text encoders. We also extract patches from the (perturbed) source image and compute their embeddings (\cref{eqn:global-local}). Cosine similarities are computed between the source embeddings and each target embedding, and the perturbation maximizes these. Optimizing against an ensemble pushes the resulting immunized image towards an unsafe region across different representation spaces, falsely causing proprietary image-editing systems to refuse to apply \textit{any} requested edit.}
\label{fig:immunization}
\end{figure*}

\methodname is motivated by the observation that black-box proprietary image-editors are equipped with a safety moderator. Image immunization by falsely triggering such a moderator is advantageous: it is easier to design the perturbation and perturbation designed for a white-box surrogate moderator readily generalizes to a black-box moderator. We explain the details of how to design such  perturbations.

\subsection{Motivation}
As discussed in \Cref{sec:related}, previous image immunization techniques require the gradient $\nabla_{x}f(\cdot)$ of the image-editing model to produce an adversarial perturbation. Since we are mainly concerned with black-box image-editing systems where $f(\cdot)$ is only accessible through an API, we take an alternative approach. %
The key observation is that most proprietary image-editing systems have a policy on  acceptable usage~\cite{xai_acceptable_use_policy_2025, google_gemini_api_terms_2026, openai_usage_policies_2025}. These policies prohibit, e.g., editing or generating copyrighted material, sexually explicit images, and images depicting graphic violence or gore. Such policies are enforced using a content moderation model $\modfilter(\cdot)$ which operates both on the pre-generation, input image $x$ to the system and the output $f(x)$ of the system, and blocks the output if the moderation score is high on either of them. Our main innovation is to use this moderation to our advantage. 
Our image immunization method, \methodname, produces perturbed images $\hat{x}$ such that $\modfilter(\hat{x})$ is high, i.e., the content moderation system is falsely triggered on the perturbed input. This can make the \methodname-perturbed image immune to unauthorized manipulation by any image-editing system with safety moderation.

An astute reader might notice that we are kicking the can down the road; the moderation model $\modfilter$ is also typically a black-box model, and producing a perturbed image $\hat{x}$ which can maximize $\modfilter(\hat{x})$ is also a challenging task to do efficiently. However, notice that this is akin to finding a transferable adversarial example for a classification task, where the target class corresponds to concepts which are prohibited in the terms-of-use of the black-box APIs. Similar problems have been well-studied in the literature, and prior work has shown that using an ensemble of strong (white-box) classifiers during search can significantly boost the transferability of adversarial examples to black-box classifiers~\cite{chen2024rethinkingmodelensembletransferbased}. Fortunately, it is easy to obtain many such strong, diverse, general-purpose image classifiers available publicly. For example, one can simply pick a self-supervised image-text encoder and convert it into a binary classifier. We note that \citet{zhang2026launderingaiauthorityadversarial} have also proposed using an ensemble of OpenCLIP~\cite{Cherti_2023} models to construct adversarial examples, but for more traditional jail-breaking tasks of misleading black-box VLMs to produce harmful images for, for example, identity manipulation and NSFW evasion.
Instead, we introduce a similar idea in \methodname, but for a different goal of  helping original content owners to freely share their creations without worrying about unauthorized, harmful manipulations. Furthermore, we note that the odds are in the favour of the immunizer here, since companies are incentivized to tolerate a higher false positive rate in $\modfilter(x)$ to defend against legal liabilities~\cite{sparks_midjourney_lawsuit_2025, gentleman_grok_xai_lawsuit_2026}.

\subsection{Constructing the immunized image}

We construct an immunized image $\hat{x} = x + \delta$ by optimizing a perturbation $\delta$ that pushes $x$ into the \emph{unsafe region} of the embedding spaces used by content-safety moderators of proprietary image-editing APIs, causing them to refuse the requested edit. We do this without assuming access to the content moderators, as we describe next.

\medskip\noindent{\textbf{Target concept set}.} We first identify a set of concepts commonly prohibited across major proprietary image-editing APIs. To do so, we conduct a systematic survey of the terms-of-use of XAI's Grok~\cite{xai_acceptable_use_policy_2025}, Google's Gemini~\cite{google_gemini_api_terms_2026}, OpenAI's GPT-Image~\cite{openai_usage_policies_2025}, and extract the concepts that are universally prohibited, including sexually explicit imagery, graphic violence, and copyrighted material. For each concept, we collect a set of publicly available images and text captions that instantiate it. We denote the resulting target set of images and texts as $\mathcal{T}$. In Figure~\ref{fig:ablation-category}, we find that sexually explicit content is more strictly moderated, and hence a better choice as a target. 

\medskip\noindent{\textbf{Surrogate embedding models}.} Content-safety filters in production APIs are known to rely on semantic embeddings to detect policy violating content. 
We therefore harness an ensemble of $n$ open-source embedding models $\{\surrmodel_i\}_{i=1}^n$ as surrogates. For each model $\surrmodel_i$, we compute the average cosine similarity between the embedding of the perturbed image $x+\delta$, and the embeddings of the targets in $\mathcal{T}$:

\begin{equation*}
    \phi_i(x+\delta,\, \mathcal{T}) \;=\; 
    \frac{1}{|\mathcal{T}|}\sum_{y \in \mathcal{T}} \cos\!\left(\surrmodel_i(x+\delta),\, \surrmodel_i(y)\right) \;\in\; [0,1].
\end{equation*}
Depending on the model family, targets $y \in \mathcal{T}$ can be either text embeddings or image embeddings (for CLIP-style models), or image embeddings (for self-supervised encoders such as DINOv2~\cite{oquab2024dinov2learningrobustvisual}), giving us complementary views of semantic similarity.

\medskip\noindent{\textbf{Optimization objective}.}
Given the surrogate ensemble and the target set, we solve:

\begin{equation}
\label{eqn:main-opt}
    \max_{\|\delta\|_\infty \leq B}\; 
    \sum_{i=1}^{n} \phi_i\!\left(x+\delta,\, \mathcal{T}\right),
\end{equation}
subject to an $\ell_{\infty}$ budget $B$ that keeps the perturbation imperceptible. Intuitively, by maximizing this objective, $x+\delta$ would yield an image whose representation is in an unsafe region of many surrogate models. This maximizes the likelihood of triggering refusal across content-moderation filters for unseen APIs. The full pipeline is illustrated in Figure~\ref{fig:immunization} (note that $\mathcal{T}$ could comprise of both textual and image targets).

\medskip\noindent{\textbf{Validation}.} After optimization, we optionally validate the immunized image by querying a black-box API directly, confirming that the requested edit is refused before delivering $\hat{x}$ to the user. We describe the implementation details, augmentation strategy, and the full validation protocol in the following sections.

\subsection{All the bells-and-whistles}
We use a number of practical techniques to better solve the optimization objective in \Cref{eqn:main-opt}. We empirically demonstrate the gain of some of the techniques in Figures~\ref{fig:ablation-models}--\ref{fig:ablation-filter}.

\subsubsection*{Models used} To maximize the chance of transfer to an unknown content moderation model $\modfilter$, we use a diverse set of surrogate models $\surrmodel_i$ in \Cref{eqn:main-opt}. We use 8 transformer-based CLIP-style models trained on diverse data~\cite{Cherti_2023, zhai2023sigmoidlosslanguageimage, fang2023datafilteringnetworks, gadre2023datacompsearchgenerationmultimodal}. We also use self-supervised image encoders~\cite{oquab2024dinov2learningrobustvisual} to produce more robust perturbations. From both these types of models, we obtain the embeddings of the perturbed image, as well as the embeddings of the target images and/or texts, and maximize the cosine similarity between the two. Finally, we also leverage open-source moderation models~\cite{zeng2025shieldgemma2robusttractable, chi2024llama} which return a scalar between $[0,1]$ denoting how unsafe an image is. We aim to maximize this score for our perturbed image. 
Figure~\ref{fig:ablation-models} shows how more models in the ensemble helps immunization, but also the gain diminishes as we increase the ensemble size and at larger perturbation bound.

\subsubsection*{Handling larger images} 
\begin{figure}[h]
    \centering
    \begin{subfigure}{0.48\linewidth}
        \centering
        \includegraphics[width=\linewidth]{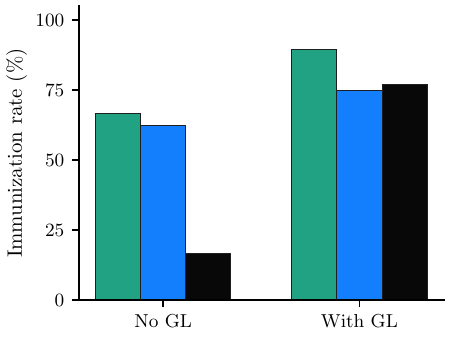}
        \caption{$||\delta||_\infty=8/255$}
    \end{subfigure}
    \hfill
    \begin{subfigure}{0.48\linewidth}
        \centering
        \includegraphics[width=\linewidth]{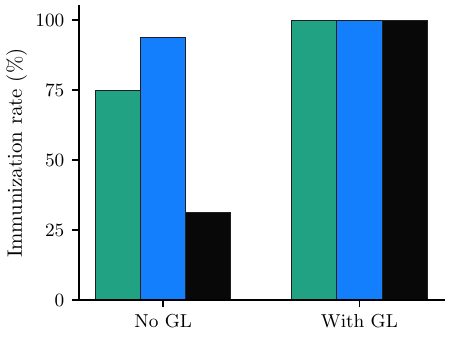}
        \caption{$||\delta||_\infty=16/255$}
    \end{subfigure}
    \caption{\textbf{Effect of global-local embeddings on immunization}. Using both global and local views in the objective in \Cref{eqn:global-local} is crucial for immunizing high resolution images.}
    \label{fig:ablation-global-local}
\end{figure}

We note that the image encoders of most embedding models have been trained on fixed resolution images of sizes less than $512\times512$ (typically $224\times224$ or $336\times336$), yet a practical immunization system must handle high-resolution images. A naive solution is to simply resize the perturbed image ($\hat{x}$) to a smaller size before passing it to the embedding models. However, this leads to spatially coarse gradients, and the full resolution perturbed image is not sufficiently immunized (see \cref{fig:ablation-global-local}). 

To address this, we decompose the optimization into \emph{global} and \emph{local} views. Let $r$ denote the encoder's native resolution. The global view ($\hat{x}_{\rm G} = \texttt{Resize}(\hat{x}, r)$) downscales the full image to $r$, capturing coarse semantic structure. The local view consists of $p$ patches $\{\hat{x}_{\rm L}^{(j)}\}_{j=1}^p$, 
each of resolution $r$, randomly cropped from $\hat{x}$ at its native resolution, preserving fine-grained detail.

For each surrogate $\phi_i$, recall from \Cref{eqn:main-opt} that $\phi_i(\cdot, \mathcal{T})$ denotes the average cosine similarity between the embeddings of the input and the embeddings of targets in $\mathcal{T}$. We extend the per-surrogate objective to incorporate both views:

\begin{equation}
\label{eqn:global-local}
\phi_i(\hat{x}_{\rm G},\, \mathcal{T}) \;+\; 
\frac{\lambda}{k}\sum_{j\in \mathrm{top}\text{-}k} 
\phi_i\!\left(\hat{x}_{\rm L}^{(j)},\, \mathcal{T}\right)
\end{equation}

\noindent where $\lambda$ is a scalar hyperparameter that balances global semantic alignment against local patch alignment, and $\mathrm{top}\text{-}k$ selects the $k \leq p$ patches most semantically aligned with $\mathcal{T}$. This empirically leads to better optimization as opposed to using the mean of all patch similarities. The full optimization objective sums \Cref{eqn:global-local} across all $n$ surrogate models, analogous to \Cref{eqn:main-opt}. We find that incorporating 
both views leads to significantly more effective immunization of high-resolution images than either view alone in 
Figure~\ref{fig:ablation-global-local}, especially  against Grok where the immunization rate increases from 17\% to 75\% at a perturbation budget of 8/255.

\subsubsection*{Augmentations and optimization} Following prior work producing robust adversarial examples~\cite{athalye2018synthesizing, zhang2026launderingaiauthorityadversarial}, we apply stochastic augmentations (augmentation $\alpha$ drawn uniformly at random from a set $\mathcal{A}$, i.e., $\alpha \sim \mathcal{A}$) to $x+\delta$ before passing it to $\phi_i$, replacing $\phi_i(x+\delta, \mathcal{T})$ with $\phi_i(\alpha(x+\delta), \mathcal{T})$ in \Cref{eqn:main-opt}. The augmentation set $\mathcal{A}$ includes random crops, resizing, gaussian blur, JPEG compression, and flips. We sample one augmentation per step, and use a straight-through gradient estimator for non-differentiable augmentations (eg. JPEG compression). We optimize \Cref{eqn:main-opt} using PGD for 5{,}000 steps with iterated FGSM~\cite{goodfellow2015explainingharnessingadversarialexamples}.

\subsubsection*{Model dropout and secant gradient caching} Optimizing \Cref{eqn:main-opt} involves computing $\nabla_\delta\, \phi_i(x+\delta, \mathcal{T})$ for each surrogate $\phi_i$ at each PGD step. This however becomes memory-intensive as the ensemble grows. To reduce the memory and computation cost of this procedure, we propose to use model dropout: at each optimization step, we select a random subset of models to compute the gradient on. To improve optimization stability for the dropped models at a step, we use an approximation of their gradients. Concretely, for each model $\phi_i$, we maintain a cache of the past two most recently computed gradients $g_i^{t-1}, g_i^{t-2}$ and the perturbations at which those gradients were computed $\delta_i^{t-1}, \delta_i^{t-2}$. 
A first order approximation of the gradient would be 
\begin{equation*}
\tilde{\nabla}_\delta\,\phi_i(\delta^t, \mathcal{T}) 
\;=\; g_i^{t-1} \;+\; 
\Bigl\langle \nabla^2\phi_i,\; 
\delta^t - \delta_i^{t-1} \Bigr\rangle.
\end{equation*}

Rather than computing the full Hessian $\nabla^2\phi_i$ at $\delta_i^{t-1}$, we use a rank-1 approximation along the secant direction $\delta_i^{t-1} - \delta_i^{t-2}$, yielding:

\[
\tilde{\nabla}_\delta\,\phi_i(\delta^t, \mathcal{T}) 
\;=\; g_i^{t-1} \;+\; 
(g_i^{t-1} - g_i^{t-2})
\frac{\langle s_i^t,\; \delta^t - \delta_i^{t-1}\rangle}
{\|s_i^t\|_2^2}\;, 
\]
where $s_i^t = \delta_i^{t-1}-\delta_i^{t-2}$. The second term is a rank-1 curvature correction. It rescales the most recent gradient change by how far the current iterate has moved along the previous secant direction. This can be viewed as a lightweight, one-secant quasi-Newton prediction of the dropped gradient, requiring only two cached gradients and their evaluation points. The approximation is inspired by classical quasi-Newton and 
secant methods, which estimate curvature from successive 
gradient differences rather than explicit Hessian 
computation~\cite{broyden1965class, nocedal1980updating, 
liu1989limited}, and is related to incremental-gradient methods 
that utilize memories of past component 
gradients~\cite{schmidt2013minimizing, defazio2014saga, 
sohl2014fast}.

\subsubsection*{Validation with a public moderation API}
For early stopping the optimization, we checkpoint $\hat{x}$ every 250 steps and could return the checkpoint with the highest score under a public black-box moderation API,~$\tilde\modfilter$, from OpenAI~\cite{openai_omni_moderation_latest} as the final immunized image. We note two important caveats. Firstly, $\tilde\modfilter$ is distinct from the internal moderation model $\modfilter$ used by commercial image-editing APIs even for OpenAI's GPT-Image: an image with high $\tilde\modfilter(\hat{x})$ can have a low 
$\modfilter(\hat{x})$ and still be edited, meaning checkpoint selection via $\tilde\modfilter$ does not guarantee transfer. 

Secondly, one might ask whether directly optimizing $\hat{x}$ to maximize $\tilde\modfilter(\hat{x})$ via black-box optimization is a stronger alternative; as we show in  %
Figure~\ref{fig:main_results}, state-of-the-art black-box attacks of SquareAttack~\cite{andriushchenko2020squareattackqueryefficientblackbox} against $\tilde{\mathcal{C}}$ produce perturbations that do not reliably transfer to the internal $\mathcal{C}$ of production APIs. In practice, we find that omitting $\tilde{\mathcal{C}}$-based checkpoint selection entirely performs comparably or better (see \Cref{fig:ablation-filter}), and we ablate this component in \Cref{sec:results}.

\section{Results}
\label{sec:results}

\begin{figure*}[t]
    \centering
    \includegraphics[width=0.9\linewidth]{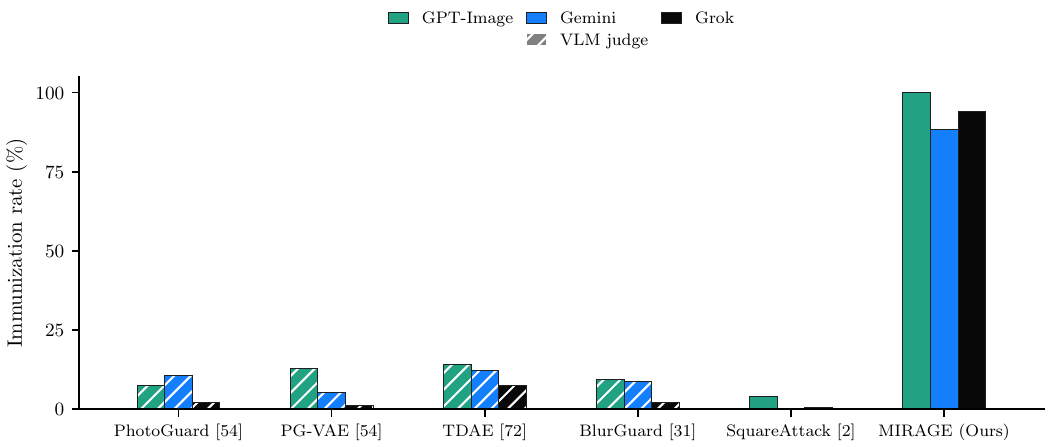}
    \caption{\textbf{Immunization rates of baselines and our method across closed-source image editing APIs (OpenAI GPT-Image, Google Gemini, XAI Grok).} We find that no existing baseline can effectively immunize against proprietary AI-powered image editors. Images immunized using \methodname  reliably trigger the moderation filters of these systems, leading to the image editing request being refused most of the time. For baseline methods which are meant to generate alternative images, the success of immunization is judged by Gemini-3.1-Flash with prompts shown in Appendix~\ref{app:judge}.}
    \label{fig:main_results}
\end{figure*}

We evaluate \methodname~against three closed-source image-editing APIs: OpenAI's GPT-Image~\cite{gpt-image-2}, Google's Gemini~\cite{gemini-3.1-flash-image}, and xAI's Grok~\cite{grok-imagine-image}, and compare against all prior immunization baselines. Our results show that existing methods fail entirely in this setting, while \methodname~achieves high immunization rates across all three APIs (\Cref{fig:main_results}). We further analyze the robustness under adversarial pressure (\Cref{sec:results-attack}). We also ablate each component of our method and analyse our design choices (\Cref{sec:results-ablations}).

\subsection{Experimental setup and baselines}

\noindent\textbf{Datasets and evaluation.} We source test images along with their prompts from ImgEditBench~\cite{ye2026imgedit}, a diverse image editing benchmark, and SHHQ~\cite{fu2022styleganhumandatacentricodysseyhuman}, a dataset of high-resolution human images with the edit prompt being \textit{``Put this person inside an office''}. Crucially, all images are evaluated at their \emph{full native resolution}---up to 1024x1024---without any downsampling prior to immunization or submission to the APIs. Notably, prior works\cite{salman2023photoguard} in immunization usually operate at a resolution of 512x512. Our main results are reported on 100 images drawn from ImgEditBench and 48 images from SHHQ; ablation experiments use the 48-image subset of SHHQ.

We report the \emph{immunization rate}: the fraction of images for which the target API's edit fails. For our method, this means that the API refuses to respond with an image, hence we measure the number of requests where a content policy violation message is returned in place of an edited image. For other baselines, since APIs might return degraded or semantically unchanged outputs rather than explicit refusal, we additionally report VLM-as-a-judge metrics~\cite{ye2026imgedit} to verify that the edit produced was incorrect. We highlight this metric in our plots wherever used. Unless otherwise specified, results are reported at a perturbation budget of $\|\delta\|_\infty \leq 16/255$.

\medskip
\noindent\textbf{APIs.} We evaluate against three production image-editing APIs: OpenAI GPT-Image-2~\cite{gpt-image-2}, Google Gemini-3.1-Flash-Image~\cite{gemini-3.1-flash-image}, and xAI Grok-Imagine-Image~\cite{grok-imagine-image}. All APIs are accessed via their standard developer endpoints. We note that these systems enforce content moderation both at the input and output stages. While \methodname targets the classifier $\modfilter$, which is triggered before any generation is returned to the user, we also report refusals which might arise because of post generation moderation. Moderation sensitivity varies across providers and can change silently over time as providers update their systems. For generating the final numbers for this paper we re-ran all our experiments on June 10, 2026. %

\medskip\noindent\textbf{Baselines.} We compare against the following immunization methods, all of which are white-box except SquareAttack:
\begin{itemize}
\item \textbf{PhotoGuard}~\cite{salman2023photoguard}: both the diffusion variant, which minimizes $\|f(x+\delta) - y_{\mathrm{tgt}}\|_2^2$, and the VAE variant, PG-VAE, which maximizes $\|\mathcal{E}(x+\delta)-\mathcal{E}(x)\|_2$ in the encoder's latent space. 
\item \textbf{BlurGuard}~\cite{kim2025blurguard}: aims to find robust and transferable perturbations by regularizing the spectrum of the noise. 
\item \textbf{TDAE}~\cite{zhang2026towards}: augments the PhotoGuard objective with a Hessian-based curvature regularizer designed to encourage transfer across diffusion model architectures.
\item \textbf{SquareAttack}~\cite{andriushchenko2020squareattackqueryefficientblackbox}: a query-efficient black-box adversarial attack applied directly to OpenAI's public content moderation API~$\tilde\modfilter$~\cite{openai_omni_moderation_latest}, representing the strongest possible direct black-box attack on an available classifier.
\end{itemize}
For all white-box baselines, we use Flux-Klein-4B~\cite{blackforestlabs_flux2_klein_2026} as the surrogate target model, as it is a state-of-the-art open-weights image editor. Hyperparameters for all methods are provided in Appendix~\ref{app:hyperparameters}, and tables containing the data of all plots can be found in App~\ref{app:tables}.

\subsection{Main results}
We present our main results in \Cref{fig:main_results}. From the plot, it is clear that \textbf{all existing baselines are ineffective in protecting images against unauthorized editing.} Notably, even baselines that explicitly target transferable immunization, such as TDAE~\cite{zhang2026towards} do not protect against image editing by proprietary models. One could contest that the low immunization rates of the baselines are a consequence of our VLM-as-a-judge setup. To address this, we also show qualitative examples of successfully and unsuccessfully immunized images (as determined by the VLM judge) in Figs \ref{fig:baseline-photoguard-qualitative-results} and \ref{fig:baseline-tdae-qualitative-results} in App~\ref{app:qualitative-baseline-results}. As we see, images are successfully immunized against the local Flux model, leading to degraded output quality. However, on the API models, the output is high quality, albeit not following the instruction in certain cases. Further, VLM-as-a-judge correctly identified the top three API edited images (rightmost column) as successfully immunized, and the bottom three as failed immunization, consistent with human judgment. This failure mode has also been acknowledged in the original results of \citet{zhang2026towards}, where weak transfer was observed when perturbations were optimized for weaker models 
(InstructPix2Pix) and transferred to stronger models (StableDiffusionv3).  %

The failure of baselines which attack the underlying image editing model motivates us to take a systems-view of this problem. Hence, we turn to attacking the content moderation API.
We find that using a black-box adversarial attack~\cite{andriushchenko2020squareattackqueryefficientblackbox} to optimize against OpenAI's public content moderation API \cite{openai_omni_moderation_latest} can lead to a non-zero refusal rate on their systems, but this does not transfer to other systems.

\begin{figure}[t]
    \centering
    \includegraphics[width=0.85\linewidth]{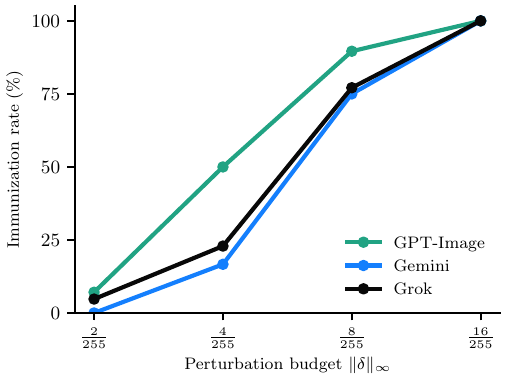}
    \caption{\textbf{Immunization vs.\ perturbation budget $\|\delta\|_\infty$}. Immunization rates for \methodname monotonically increase with increasing $\|\delta\|_\infty$ at the cost of larger visual distortions to the image. Qualitative example images of varying perturbation level are shown in Figure~\ref{fig:qualitative-results}.}
    \label{fig:perturbation_budget}
\end{figure}

On the other hand, \methodname provides almost perfect protection against GPT-Image, and immunization rates close to 90\% on Gemini and Grok. This difference is because unlike previous baselines, our method does not assume anything about the underlying model powering the image editor, and works against the content moderation of the system instead. We also show the trade-off between $||\delta||_\infty$ and immunization rate in \Cref{fig:perturbation_budget} on SHHQ data. We find that increasing the perturbation budget monotonically leads to higher immunization rates. In \Cref{fig:qualitative-results} (and \Cref{fig:qualitative-results-appendix} in the appendix) we show examples of perturbed images -- higher perturbation budgets lead to visible artifacts for more effective immunization.

\subsection{Results under attack}
\label{sec:results-attack}

\begin{figure*}[t]
    \centering
    \includegraphics[width=\textwidth]{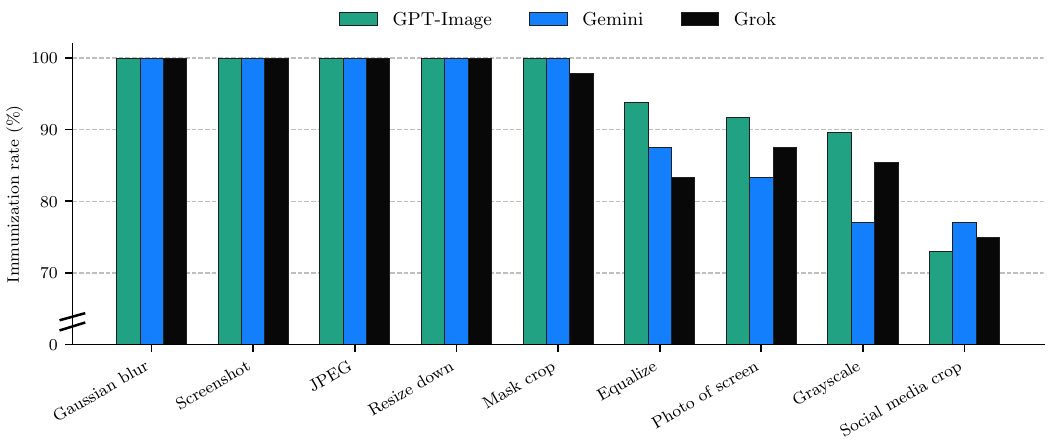}
    \caption{\textbf{Immunization rate of our method under weak adversarial perturbations}. \methodname survives most attacks under the weak threat model which restricts the adversary to using classical image pre-processing transforms.}
    \label{fig:robustness}
\end{figure*}

To test the robustness of our proposed approach,  \methodname, we measure its immunization performance on the 48-image subset of the SHHQ dataset under adversarial pressure. The adversary here is a malicious actor who wants to edit the immunized images. They can pre-process the image before sending it to the proprietary image editing AI system. As we noted in \Cref{sec:threat}, we consider two kinds of adversaries. The  \textit{weak} adversaries are restricted to using classical image transforms to bypass immunization, while the \textit{strong} adversaries can expend more computational resources and use some white-box open-source image embedding, segmentation, generation or editing models. %
 
\subsubsection{Weak adversary} 
In \Cref{fig:robustness}, we present robustness results under image transforms that are within the ambit of a weak adversary. We find that \methodname is robust against most such image transformations, with worst case robustness of 72\% for GPT-Image with social media preview style cropping. We detail each adversary below.
\begin{enumerate}[leftmargin=*]
    \item \texttt{Gaussian Blur}: This smooths the image to potentially disrupt high frequency perturbations. This is motivated by successful countermeasure to adversarial examples that tend to add high-frequency perturbations in \cite{ziyadinov2023low}. We use a radius of 1 for the Gaussian Blur. We find that \methodname survives this transformation perfectly.
    \item \texttt{JPEG compression}: The adversary compresses the immunized image using JPEG, reducing quality to 50\%. This can induce artifacts into the image, however, our immunization scheme is unaffected by this.
    \item \texttt{Resize down}: The adversary resizes the image to 50\% of its original size, potentially erasing fine-grained perturbations. However, \methodname sees no degradation in immunization rates under this.
    \item \texttt{Screenshot}: We simulate an adversary taking a screenshot of the immunized image by resizing the long side to 1080, applying a slight Gaussian blur, then JPEG compression with quality 75. This also does not lead to a degradation in immunization rates.
    \item \texttt{Subject Crop}: An adversary could crop the image to only keep the foreground subject and eliminate any perturbations contained fully in the background. We obtain the fore-ground bounding boxes from the ground truth SHHQ annotations. We find a small drop in immunization rates for Grok here.
    \item \texttt{Equalize}: The adversary can perform histogram normalization to remove immunization artifacts. We find that this can reduce the effectiveness of \methodname, but the overall rate remains high (81\% for Grok).
    \item \texttt{Photo of a screen}: An adversary can take a photo of a screen displaying the immunized image, in the hopes that this strips away the immunization. We simulate this by noising, blurring, rotating, and increasing the brightness of the image. This degrades the immunization mildly, though the overall rate remains higher than 80\%.
    \item \texttt{Greyscale}: The adversary could convert the image to greyscale to strip away perturbations that depend on colors. We find that this reduces the immunization rate of \methodname mildly, with performance on Gemini dropping to close to 79\%. 
    \item \texttt{Social Media Crop}: Social media platforms often crop the preview image to be square. We simulate this by cropping a square around the main subject of the image, followed by adjusting the contrast and sharpening the image. This leads to a moderate drop in the immunization rates of \methodname, dropping to about 72\% for GPT-Image. 
\end{enumerate}

\subsubsection{Strong adversary} 
\begin{figure}[t]
    \centering
    \includegraphics[width=0.9\linewidth]{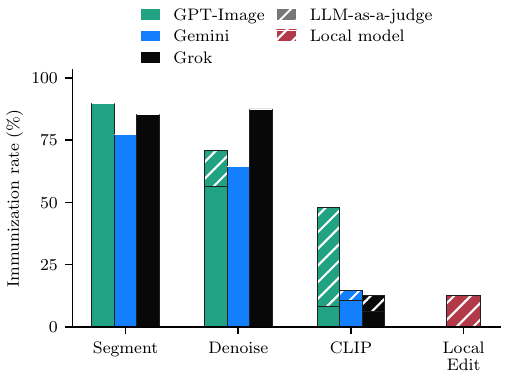}
    \caption{\textbf{Immunization rate of our method under stronger adversarial perturbations}. If the adversary can run embedding, generation or segmentation models, they can construct powerful perturbations to bypass the immunization afforded by \methodname. VLM-as-a-judge shows the proportion of images that passes the moderation filter but resulted in significantly distorted output. Local model shows immunization rate against a local diffusion model.}
    \label{fig:strong-robustness}
\end{figure}

In \Cref{fig:strong-robustness}, we present results under a strong adversary. We consider four adaptive attacks under this threat model. 
\begin{enumerate}[leftmargin=*]
    \item \texttt{Segment}: An adversary could isolate the foreground in a more fine-grained manner as compared to the \texttt{Mask Crop} attack in Table~\ref{tab:attack_results}. They can segment the image with an off-the-shelf model, mask the foreground precisely and paste it onto a white background. We find that this is moderately successful against \methodname, leading to immunization rates under 80\% for Gemini.
    \item \texttt{Denoise}: This adversary first uses a local diffusion-based image editing model (Flux-v.2-Klein-4B) to denoise or ``purify"\cite{nie2022DiffPure} the image. Instead of using the classical DiffPure approach, we adapt the attack from \citet{pleimling2026offtheshelfimagetoimagemodelsneed}, where the adversary first prompts the local model to denoise the image, and then passes the ``denoised" image to the black-box API. This is done in a two stage manner---denoising followed by a call to a proprietary model---since the local model might not be as performant as the proprietary model, and could give degraded outputs when tasked with the actual malicious image edit. We find that this is moderately successful in erasing the immunization, with about 56\% refusal rate on GPT-Image after purification. However, taking into account the quality of the image edited by the proprietary systems, we find that the worst immunization rate under attack is 65\% for Gemini.
    \item \texttt{CLIP}: An adversary could try to undo the immunization by minimizing the objective in \Cref{eqn:main-opt} using a similar set of public embedding models. We simulate such an adversary by choosing a subset of 4 embedding models, and minimizing the similarity from the target concepts. We run PGD for 2000 steps, with a perturbation bound $||\delta||_\infty \leq 16/255$. This is lower than the resources spent by the defender. However, we find that this attack is pretty successful, reducing the refusal rate to 12.5\% for Gemini and Grok.     
    On the other hand, %
     This leads to some degradation in the image, and the VLM-as-a-judge says that the two people are different.

    \item \texttt{Local Edit}: This is the most potent attack that an adversary can launch, since it completely bypasses the moderation pathway that \methodname exploits. Indeed, as we see in the plot this attack can lead to very low immunization rates. However, we note that if the image owner knew what local diffusion model was used by the adversary, they could add a Photoguard style objective to \Cref{eqn:main-opt} and provide some protection. We describe preliminary results of such compound immunization in App~\ref{app:mirage-with-pg}, and find this to be a promising direction. %
\end{enumerate}

\subsection{Analysis : What is essential for immunization?}
\label{sec:results-ablations}

In this section, we perform an ablation analysis on each component of \methodname to understand the contribution of each component to the immunization performance. These experiments are performed on the 48-image subset of SHHQ~\cite{fu2022styleganhumandatacentricodysseyhuman} which primarily consists of images of people. We measure immunization across two perturbation norms - $||\delta||_\infty \leq 8/255$ and $||\delta||_\infty \leq 16/255$. \methodname is tested on three AI-powered proprietary image-editors: GPT-Image, Gemini, and Grok.

\subsubsection{Ensemble size}

\begin{figure}[h]
    \centering
    \begin{subfigure}{0.48\linewidth}
        \centering
        \includegraphics[width=\linewidth]{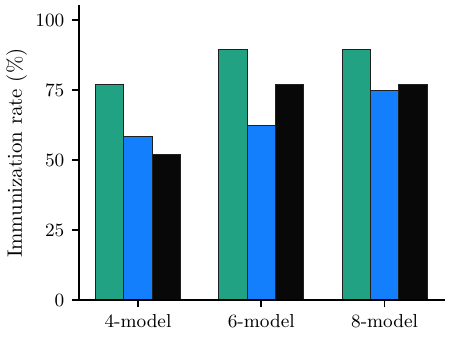}
        \caption{$||\delta||_\infty=8/255$}
    \end{subfigure}
    \hfill
    \begin{subfigure}{0.48\linewidth}
        \centering
        \includegraphics[width=\linewidth]{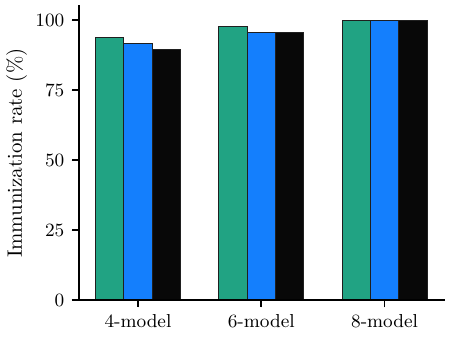}
        \caption{$||\delta||_\infty=16/255$}
    \end{subfigure}
    \caption{{\bf Effect of ensemble size on immunization rate.} More models $n\in\{4,6,8\}$ in the ensemble objective Eq.~\eqref{eqn:main-opt} makes the immunization generalize better, resulting in improved success rates, but the gain diminishes with larger ensemble and higher perturbation bound.
    }
    \label{fig:ablation-models}
\end{figure}
Our target for optimization in \Cref{eqn:main-opt} is over an ensemble of models. For our main experiments, we choose an ensemble of size 8. In \Cref{fig:ablation-models} we investigate how immunization changes with different values of the ensemble size. Across these experiments, we ensure that we cover similar types of models. The exact list of models can be found in Table~\ref{tab:mirage_hyperparameters}. As depicted in \Cref{fig:ablation-models}, optimizing against a larger surrogate ensemble monotonically leads to better immunization rates. This effect is starker at smaller bounds, where larger ensembles lead to more generalizable perturbations which can mislead the content moderation systems of the proprietary APIs.

\subsubsection{Global-Local (GL) image views}

In order to deal with high resolution input images, we introduce a mechanism to construct global and local image embeddings according to \Cref{eqn:global-local}. For \methodname (labeled With GL in the plots), we use 16 random patches to construct the local view, and use top-k as the aggregation function. For the No GL baseline, we simply resize the perturbed image before passing it to each surrogate model $\surrmodel_i$ to compute the loss. As we see from the plot, using global and local views of the image greatly boosts the immunization rates. This is critical for immunizing against Grok, where the immunization rate goes from 17\% to 75\% at a perturbation budget of 8, making the method practical.

\subsubsection{Targets chosen}

\begin{figure}[h]
    \centering
    \begin{subfigure}{0.48\linewidth}
        \centering
        \includegraphics[width=\linewidth]{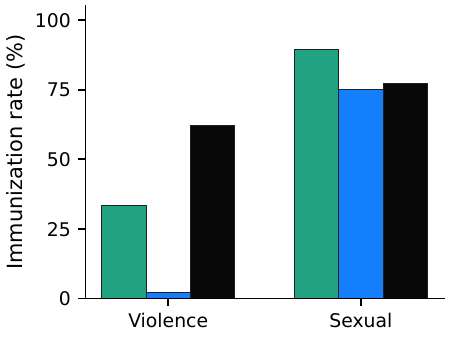}
        \caption{$||\delta||_\infty=8/255$}
    \end{subfigure}
    \hfill
    \begin{subfigure}{0.48\linewidth}
        \centering
        \includegraphics[width=\linewidth]{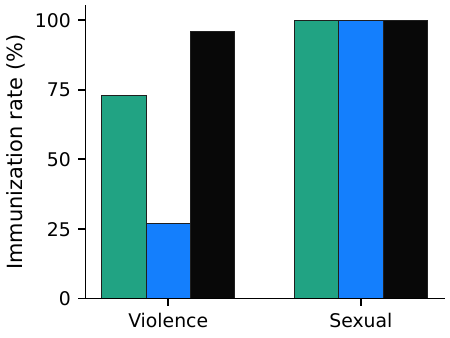}
        \caption{$||\delta||_\infty=16/255$}
    \end{subfigure}
    \caption{\textbf{Effect of target category on immunization rate}. Using sexually explicit content as the target concept leads to higher refusals as compared to violent imagery.}
    \label{fig:ablation-category}
\end{figure}
For our main experiments, we choose the targets %
to be a set $\mathcal{T}$ of 5 images and 3 text prompts corresponding to sexually explicit content. In \Cref{fig:ablation-category} we also experiment with using the same number of images and texts corresponding to graphic violence or gore. As we see, violence is less strictly moderated for GPT-Image and Gemini, leading to lower refusals and immunization rates. However, the content filter for Grok is much more strict for graphic violence or gore, and hence we see a higher immunization rate across perturbation bounds.

\subsubsection{Using $\tilde\modfilter$}
\begin{figure}
    \centering
    \begin{subfigure}{0.48\linewidth}
        \centering
        \includegraphics[width=\linewidth]{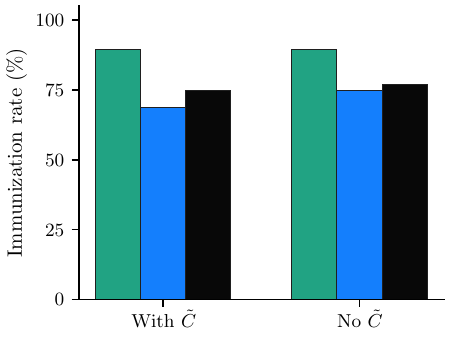}
        \caption{$||\delta||_\infty=8/255$}
    \end{subfigure}
    \hfill
    \begin{subfigure}{0.48\linewidth}
        \centering
        \includegraphics[width=\linewidth]{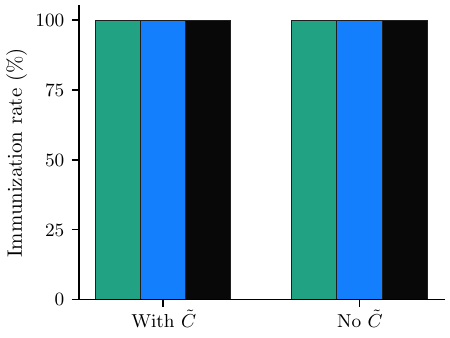}
        \caption{$||\delta||_\infty=16/255$}
    \end{subfigure}
    \caption{\textbf{Effect of using the surrogate moderation objective for validation}. We find that using a public $\tilde\modfilter$ from OpenAI to early stop the optimization can lead to worse immunization on non-OpenAI models (blue and black).}
    \label{fig:ablation-filter}
\end{figure}

Since a public content moderation API is available from OpenAI \cite{openai_omni_moderation_latest}, one could use it to early stop the immunization procedure. However, this leads to poorer generalization across models, as seen in \Cref{fig:ablation-filter}.

\begin{figure*}[t]
    \centering
    \begin{subfigure}{0.24\textwidth}
        \centering
        \includegraphics[width=0.48\linewidth]{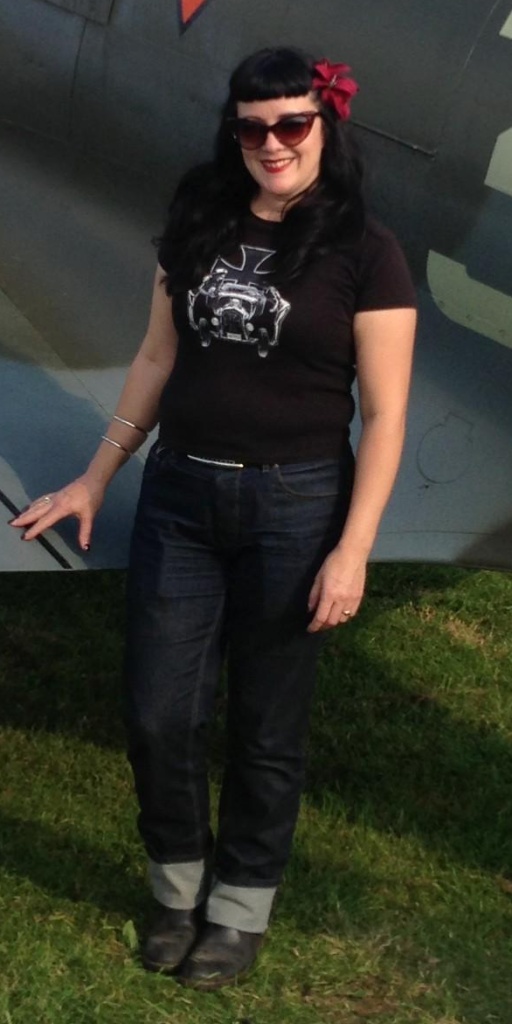}
        \includegraphics[width=0.48\linewidth]{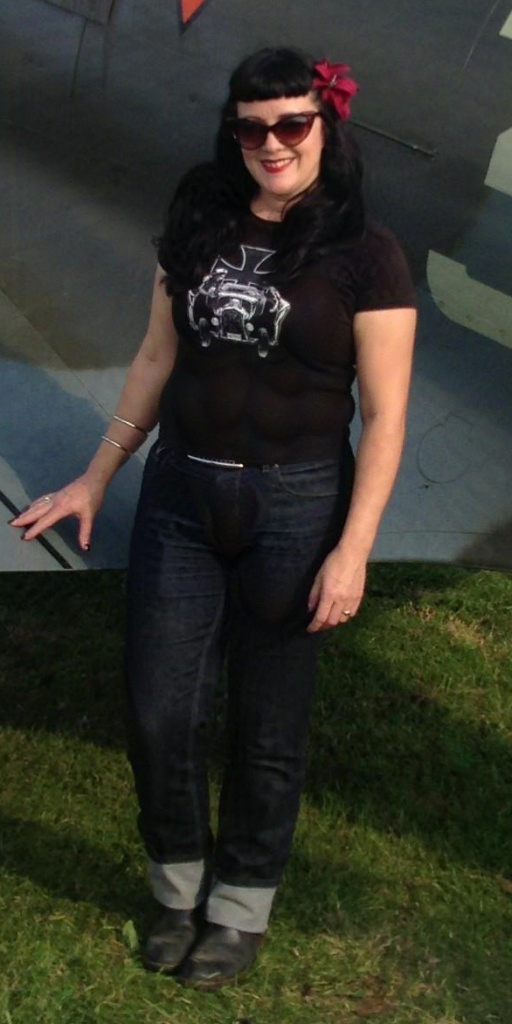}
        \caption{$||\delta||_\infty=2$}
    \end{subfigure}\hfill
    \begin{subfigure}{0.24\textwidth}
        \centering
        \includegraphics[width=0.48\linewidth]{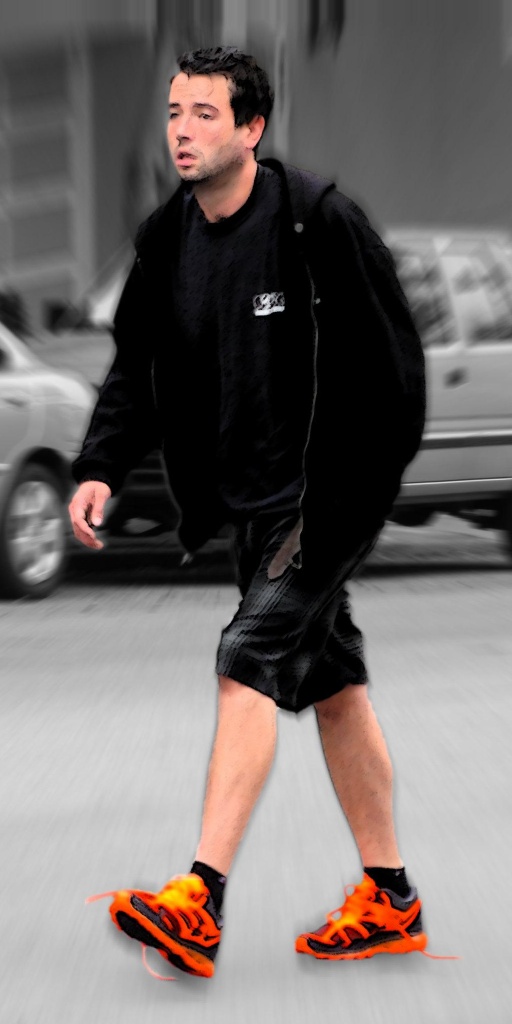}
        \includegraphics[width=0.48\linewidth]{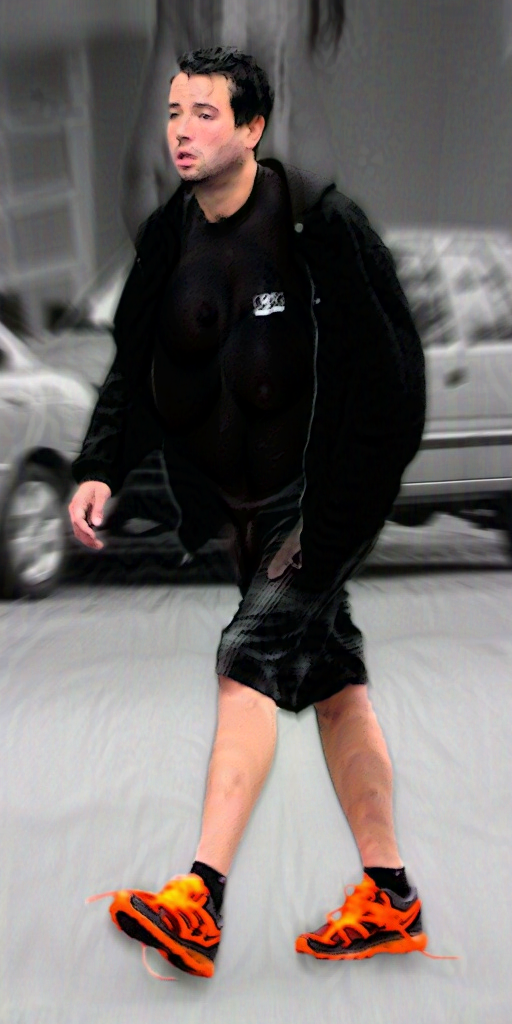}
        \caption{$||\delta||_\infty=4$}
    \end{subfigure}\hfill
    \begin{subfigure}{0.24\textwidth}
        \centering
        \includegraphics[width=0.48\linewidth]{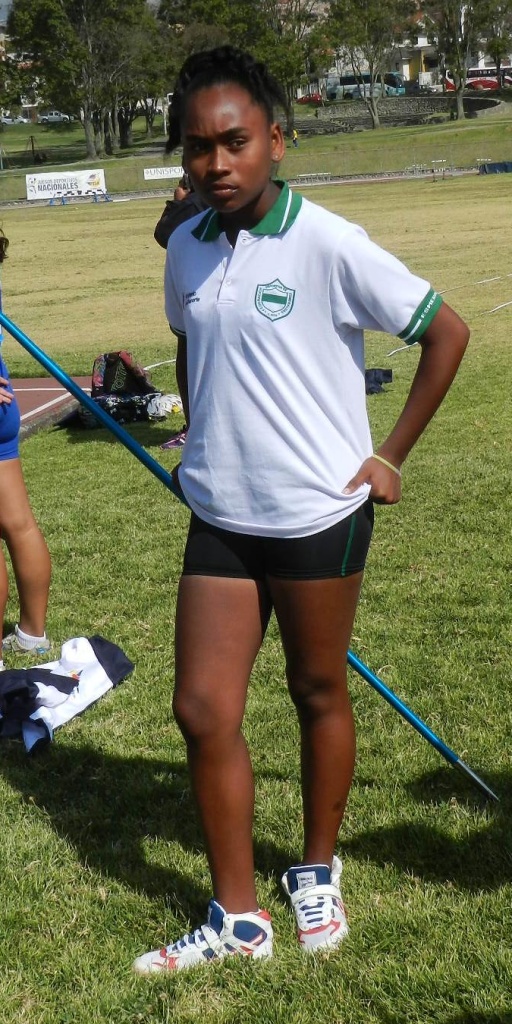}
        \includegraphics[width=0.48\linewidth]{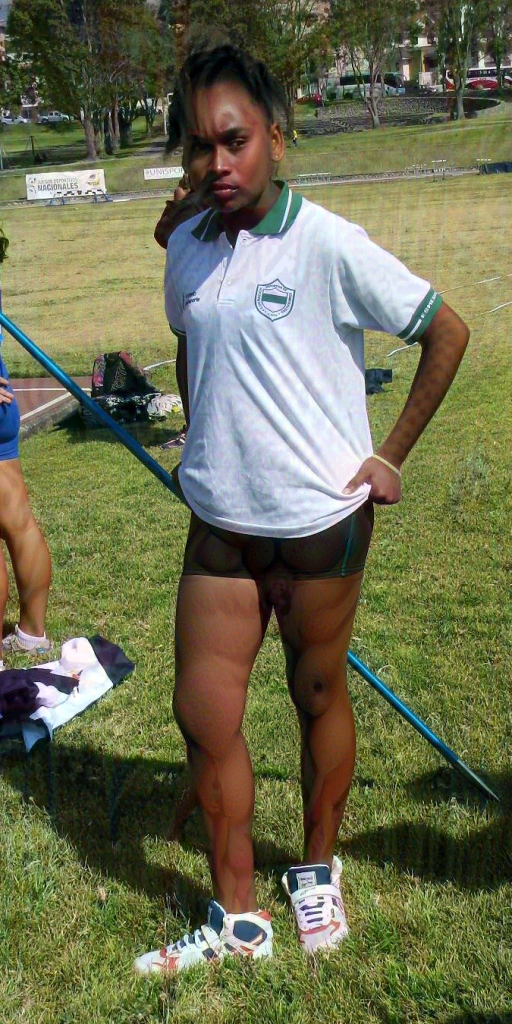}
        \caption{$||\delta||_\infty=8$}
    \end{subfigure}\hfill
    \begin{subfigure}{0.24\textwidth}
        \centering
        \includegraphics[width=0.48\linewidth]{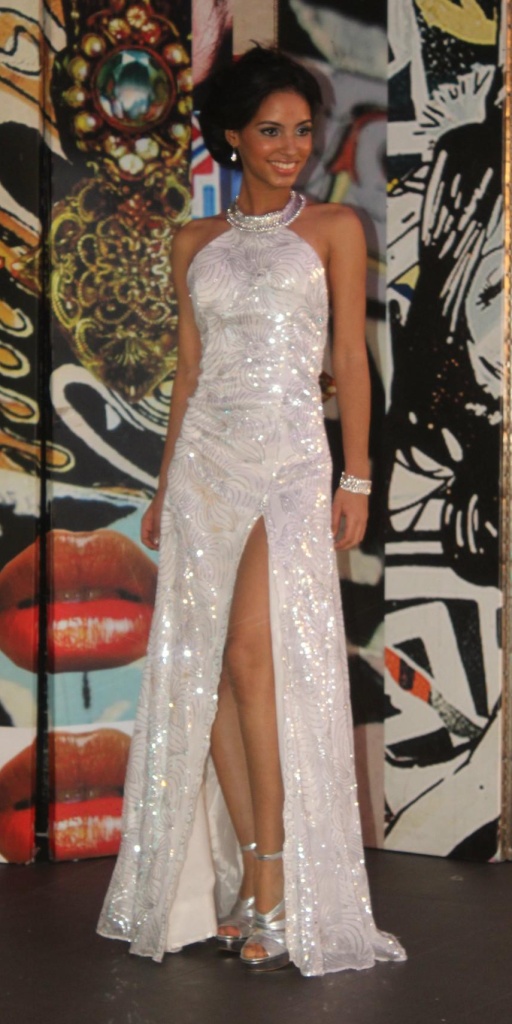}
        \includegraphics[width=0.48\linewidth]{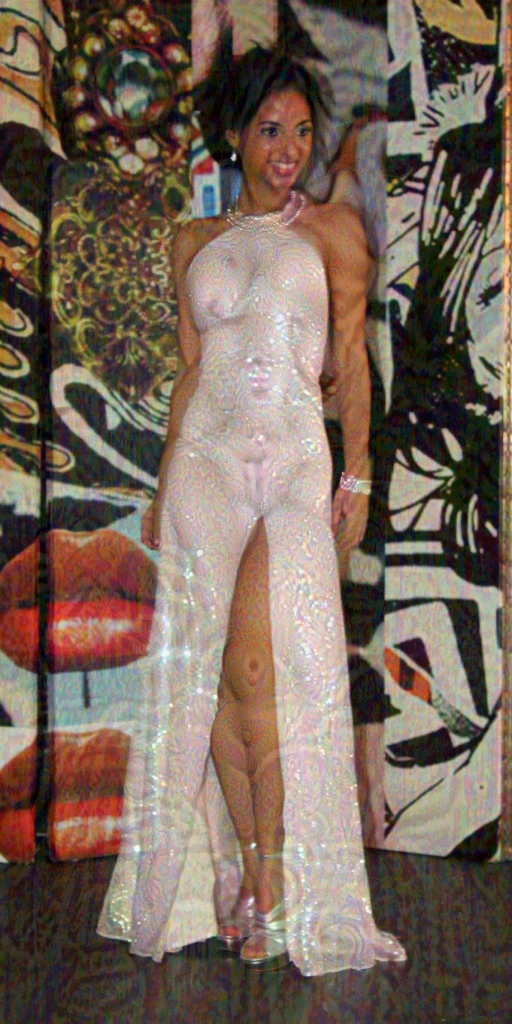}
        \caption{$||\delta||_\infty=16$}
    \end{subfigure}
    \caption{Qualitative examples of original source images and their successfully immunized counterparts at different perturbation budgets.}
    \label{fig:qualitative-results}
\end{figure*}

\section{Discussion}
\subsubsection*{\textbf{A new defense surface for image immunization}}
Our work demonstrates that the content moderation pipeline common to all major image-editing APIs constitutes a practically effective and under-exploited defense surface for image immunization. Rather than attacking the generative model directly we exploit the simpler safety classifiers that gate these systems. This reframing has a key practical advantage: while generative pipelines might differ substantially across providers in their architecture, preprocessing or data, moderation filters seem to have common characteristics due to a shared objective of detecting policy violating content, making transfer-based attacks significantly more tractable. Our results confirm this, achieving up to 90\% ASR on GPT-Image~\cite{gpt-image-2}, Google Gemini~\cite{gemini-3.1-flash-image}, and Grok~\cite{grok-imagine-image} while all prior immunization baselines which rely on attacking diffusion models achieve near zero transfer rates.

\subsubsection*{\textbf{Prompt agnostic protection}}
A distinctive property of false moderation is that it is inherently prompt-agnostic: once an image is immunized, it is protected against \emph{all} editing instructions. Prior white-box methods that perturb images to disrupt specific generative pathways offer no such guarantee, since an adversary can trivially switch prompts or models. Our approach eliminates this vulnerability by targeting the moderation bottleneck upstream of generation. 

\subsubsection*{\textbf{The asymmetric advantage of the defender}}
We note a structural asymmetry that benefits the image owners. Commercial providers are strongly incentivized to maintain low false-negative rates in their moderation system, since the legal and reputational costs of generating harmful content far outweigh the cost of occasionally refusing a benign edit~\cite{sparks_midjourney_lawsuit_2025,gentleman_grok_xai_lawsuit_2026}. This means that the moderation decision boundary is over-conservative towards refusal, and our immunized images need only push into a region of embedding space where false positives are tolerated.
This advantage makes false moderation viable even under the challenging black-box transfer-only regime. 

\subsubsection*{\textbf{Limitations}}
Our method has several limitations that future works could address. Firstly, the perturbation budget $\|\delta\|_{\infty}$ required for reliable immunization across APIs can possibly be reduced through more sophisticated optimization techniques for better visual fidelity of immunized images. Secondly, while we demonstrate robustness against a range of weak adversarial transformations, a sufficiently strong adaptive adversary with access to open-source CLIP models or local diffusion purification can circumvent immunization. This suggests that robustifying against adaptive adversaries through adversarial training of perturbation optimization or min-max formulations remains an open problem. Thirdly, our evaluation focuses on live commercial APIs; as providers update their moderation systems, immunization effectiveness may change. Notably, we observed temporal variability in the moderation behavior of certain APIs; for instance, attack success rate against Grok fluctuated significantly between early and mid-June, suggesting that black-box APIs could undergo silent updates that affect their robustness characteristics.

\subsubsection*{\textbf{Future Directions}}
Several natural extensions follow from this work. Combining the false moderation objective with a PhotoGuard-style latent disruption term could yield immunization effective against both black-box commercial APIs and white-box open-source editors simultaneously, and warrants deeper investigation. More broadly, our approach raises the question of whether other shared structural properties of commercial AI systems such as watermark detectors, copyright filters, or output safety classifiers can be similarly exploited for user-side protection. Finally, as multimodal models increasingly operate on video and audio in addition to images, extending false moderation to these modalities is an important direction for protecting individuals in a wider range of consent-violation scenarios.

\section{Conclusion}
Prior image immunization methods have implicitly assumed that the defender must attack the same generative machinery the adversary uses. This assumption breaks down in the black-box setting where the proprietary image-editing models are architecturally diverse, rapidly updated and entirely inaccessible to gradient-based optimization. As a result, all existing immunization methods fail against the commercial APIs that pose the greatest threat to independent content generators. %

We identified a structural property that cuts across all major commercial image-editing systems regardless of their generative architecture: a safety classifier whose refusal is prompt-agnostic and unconditional. By targeting this shared bottleneck rather than the generative model itself, we sidestep the transferability problem that has plagued prior works. Our method, false moderation, optimizes adversarial perturbations to align the immunized image with policy-violating concepts in the embedding space of an open-source surrogate ensemble, causing commercial APIs to refuse to produce any edit of the protected image.

Empirically, false moderation achieves close to 90\% attack success rate on GPT-Image, Gemini and Grok APIs on which every prior immunization baseline achieves very low rates and remains largely effective under a range of adversarial image transformations applied by a weak adversary. The perturbations are visually imperceptible, prompt-agnostic, and require no knowledge of the target system beyond its publicly-stated usage policy.

We hope this work shifts the framing of image immunization research: rather than an arms race against individual generative architectures, responsible use-policies and their enforcement mechanisms represent a more stable and universally shared attack surface for defenders. As AI-powered image manipulation becomes increasingly accessible, practical tools that protect individuals from non-consensual editing without requiring white-box model access will become correspondingly more important. 

\section*{Acknowledgement}
This work is in part supported by NSF grants 2112471, 2229876, 2505865, and 2502281.

\bibliographystyle{abbrvnat}
\bibliography{references_conf}

\clearpage
\appendices

\section{Additional Results}
\label{app:tables}
We provide tabular versions of all figures in the main paper for reference and reproducibility.

\Cref{tab:baselines} reports the main immunization rates for all baselines and \methodname~across the three target APIs. Baseline entries other than SquareAttack and \methodname~are VLM-as-a-judge results, since white-box methods do not produce explicit API refusals but may degrade edit quality. \Cref{tab:perturbation_budget} reports immunization rates as a function of perturbation budget $\|\delta\|_\infty$ for \methodname. \Cref{tab:attack_results} reports immunization rates under all weak, compound, and strong adversarial transformations; for strong attacks, parenthesized values are VLM-as-a-judge rates where available. \Cref{tab:ablation_refusals} reports the full ablation results on the 48-image SHHQ subset at both perturbation budgets.

\begin{table}[!ht]
    \centering
    \scriptsize
    \begin{tabular}{lccc}
        \toprule
        Method & GPT-Image & Gemini & Grok \\
        \midrule
        PhotoGuard & 7.4\% & 10.8\% & 2.0\% \\
        PhotoGuard-VAE & 12.8\% & 5.4\% & 2.0\% \\
        TDAE & 14.2\% & 12.2\% & 7.4\% \\
        BlurGuard & 9.5\% & 8.8\% & 2.0\% \\
        SquareAttack & 4.1\% & 0.0\% & 0.7\% \\
        Ours & 100.0\% & 88.5\% & 94.0\% \\
        \bottomrule
    \end{tabular}
    \caption{Main immunization rates for baselines and \methodname. Baseline entries other than SquareAttack and \methodname are VLM-as-a-judge results.}
    \label{tab:baselines}
\end{table}

\begin{table}[!ht]
    \centering
    \scriptsize
    \begin{tabular}{lcccccc}
        \toprule
        Ablation & \multicolumn{3}{c}{$||\delta||_\infty=8$} & \multicolumn{3}{c}{$||\delta||_\infty=16$} \\
        \cmidrule(lr){2-4}\cmidrule(lr){5-7}
         & OAI & Gemini & Grok & OAI & Gemini & Grok \\
        \midrule
        No Global-local & 66.7\% & 62.5\% & 16.7\% & 75.0\% & 93.8\% & 31.2\% \\
        6-model & 89.6\% & 62.5\% & 77.1\% & 97.9\% & 95.8\% & 95.8\% \\
        4-model & 77.1\% & 58.3\% & 52.1\% & 93.8\% & 91.7\% & 89.6\% \\
        Violence target & 33.3\% & 2.1\% & 62.1\% & 72.9\% & 27.1\% & 95.8\% \\
        No $\tilde\modfilter$ & 89.6\% & 75.0\% & 77.1\% & 100.0\% & 100.0\% & 100.0\% \\
        Full & 89.6\% & 68.8\% & 75.0\% & 100.0\% & 100.0\% & 100.0\% \\
        \bottomrule
    \end{tabular}
    \caption{Immunization rates for SHHQ ablations.}
    \label{tab:ablation_refusals}
\end{table}

\begin{table}
    \centering
    \scriptsize
    \resizebox{\linewidth}{!}{%
    \begin{tabular}{llccc}
        \toprule
        Attack & Type & GPT-Image & Gemini & Grok \\
        \midrule
        Mask crop & Weak & 100.0\% & 100.0\% & 97.9\% \\
        Equalize & Weak & 93.8\% & 87.5\% & 83.3\% \\
        Gaussian blur & Weak & 100.0\% & 100.0\% & 100.0\% \\
        Grayscale & Weak & 89.6\% & 77.1\% & 85.4\% \\
        JPEG & Weak & 100.0\% & 100.0\% & 100.0\% \\
        Resize down & Weak & 100.0\% & 100.0\% & 100.0\% \\
        Screenshot & Compound & 100.0\% & 100.0\% & 100.0\% \\
        Photo of screen & Compound & 91.7\% & 83.3\% & 87.5\% \\
        Social media crop & Compound & 73.0\% & 77.0\% & 75.0\% \\
        Foreground white & Strong & 89.6\% (89.6\%) & 77.1\% (77.1\%) & 85.4\% (85.4\%) \\
        Diffusion denoising & Strong & 56.2\% (70.8\%) & 64.6\% (64.6\%) & 78.0\% (78.0\%) \\
        CLIP attack & Strong & 8.3\% (47.9\%) & 10.5\% (14.6\%) & 6.2\% (12.5\%) \\
        \bottomrule
    \end{tabular}}
    \caption{Immunization rates under adversarial transformations. For strong attacks, parenthesized values are VLM-as-a-judge immunization rates where available.}
    \label{tab:attack_results}
\end{table}

\begin{table}
    \centering
    \scriptsize
    \begin{tabular}{lccc}
        \toprule
        $\|\delta\|_\infty$ & GPT-Image & Gemini & Grok \\
        \midrule
        $2/255$ & 7.1\% & 0.0\% & 4.8\% \\
        $4/255$ & 50.0\% & 16.7\% & 22.9\% \\
        $8/255$ & 89.6\% & 75.0\% & 77.1\% \\
        $16/255$ & 100.0\% & 100.0\% & 100.0\% \\
        \bottomrule
    \end{tabular}
    \caption{Perturbation budget versus immunization rate for \methodname.}
    \label{tab:perturbation_budget}
\end{table}

\section{Hyper-parameters and Implementation Details}
\label{app:hyperparameters}

\Cref{tab:mirage_hyperparameters} lists the full set of 
hyperparameters used for \methodname in all main 
experiments. We use PGD with a cosine step-size schedule 
for 5{,}000 steps. The surrogate ensemble consists of 8 
models spanning CLIP-style vision-language encoders, a 
self-supervised image encoder (DINOv2), and an open-source 
moderation model (ShieldGemma-2). Global-local optimization 
uses 16 randomly sampled patches of resolution $255 \times 255$ at the image's native resolution, with top-$k$ mean aggregation ($k=8$) over patch similarities. Model dropout is applied with probability 0.3, with a minimum of 3 surrogates selected per step.

\begin{table}[!ht]
    \centering
    \scriptsize
    \begin{tabular}{p{0.35\columnwidth}p{0.55\columnwidth}}
        \toprule
        Parameter & Value \\
        \midrule
        Attack & $\ell_\infty$ PGD \\
        Steps & 5000 \\
        Step size & 1.0 \\
        Schedule & Cosine \\
        Objective category & Sexual \\
        EOT samples & 1 \\
        Patch size & 256 \\
        Number of local patches & 16 \\
        Patch sampling mode & Random \\
        Local view weight & 0.25 \\
        Local pooling & Top-$k$ mean \\
        Local pooling $k$ & 8 \\
        Surrogate drop probability & 0.3 \\
        Minimum selected surrogates & 3 \\
        Surrogate ensemble &
        \url{hf_dinov2:facebook/dinov2-base};
        \url{hf_siglip:google/siglip-base-patch16-224};
        \url{open_clip:ViT-B-32/laion2b_s34b_b79k};
        \url{open_clip:ViT-B-16/laion2b_s34b_b88k};
        \url{open_clip:ViT-L-14/datacomp_xl_s13b_b90k};
        \url{open_clip:ViT-L-14/dfn2b_s39b};
        \url{open_clip:ViT-H-14/dfn5b};
        \url{shieldgemma:google/shieldgemma-2-4b-it:sexual} \\
        \bottomrule
    \end{tabular}
    \caption{Hyper-parameters for \methodname used in the main experiments.}
    \label{tab:mirage_hyperparameters}
\end{table}
It takes 15 minutes to immunize a single 1024x1024 image on 1 A40 GPU if we use 5000 steps.
\subsection{Baselines}

\Cref{tab:baseline_hyperparameters} lists the hyperparameters used for all baseline methods. For all white-box baselines, we use Flux-Klein 4B~\cite{blackforestlabs_flux2_klein_2026} 
as the surrogate target model. All methods are run at 
$\|\delta\|_\infty \leq 16/255$ unless otherwise noted.

\begin{table}[!ht]
    \centering
    \scriptsize
    \begin{tabular}{p{0.34\columnwidth}p{0.56\columnwidth}}
        \toprule
        Method & Hyper-parameters \\
        \midrule
        SquareAttack & Score loss, stripe initialization, random start, $\epsilon=16/255$, 500 steps, sexual category, threshold 0.5, no early stopping, $p_{\mathrm{init}}=0.01$, no max-side resizing \\
        \midrule
        BlurGuard & $\epsilon=16/255$, 200 steps, step size 1.0, learning rate 0.06, $\epsilon_\sigma=0.0$, $\sigma$ weighting 10.0, 50 warmup steps, seed 0 \\
        \midrule
        PhotoGuard-VAE & $\epsilon=16/255$, 200 iterations, step size 1.0, 1 gradient repetition, $\ell_\infty$ norm, constant schedule, minimum step-size factor 0.0 \\
        \midrule
        PhotoGuard-Diffusion & $\epsilon=16/255$, 200 iterations, step size 1.0, 5 gradient repetitions, zero image target, $\ell_\infty$ norm, constant schedule, minimum step-size factor 0.0 \\
        \midrule
        TDAE & $\epsilon=16/255$, 200 iterations, step size 1.0, zero image target, Hessian step size 2.0, curvature regularizer 1.0, $\ell_\infty$ norm, constant schedule, minimum step-size factor 0.0 \\
        \bottomrule
        
    \end{tabular}
    \caption{Hyper-parameters for baseline methods.}
    \label{tab:baseline_hyperparameters}
\end{table}

\subsection{Judge prompts}
\label{app:judge} 

We use Gemini-Flash-3.1 as our judge for edits. We use the following judging prompts ---

\subsubsection*{SHHQ} 
``You are judging whether two images depict the same person.

Compare the person's visible identity cues, including face shape, facial features, hairline, hair style, skin tone, and any stable identifying details. Ignore pose, crop, lighting, background, image quality, clothing changes, and style changes when they are not identity cues.

Return only a JSON object with:
- same-person: true if Image B appears to depict the same person as Image A, false if it appears to depict a different person, null if impossible to determine.
- confidence: a number from 0 to 1.
- rationale: one concise sentence."

\subsubsection*{ImgEdit}
``You are judging whether an image edit succeeded.

You will see Image A before editing, Image B after editing, and the edit instruction. Evaluate whether:
1. The edit instruction was followed very faithfully.
2. No extraneous edit was added.
3. The original subject and content that should not change were preserved.
4. Visual quality was preserved.

Return only a JSON object with:
- instruction-followed: true or false.
- no-extraneous-edits: true or false.
- subject-preserved: true or false.
- quality-preserved: true or false.
- overall-success: true only if all four criteria are true, otherwise false.
- score: an integer from 1 to 5, where 5 is near-perfect, 4 has only minor issues, and 3 or lower has meaningful failures.
- rationale: one concise sentence of no more than 25 words."

\section{Additional Qualitative Results}
\label{app:qualitative}

\begin{figure}[t]
    \centering

    \begin{subfigure}{0.48\linewidth}
        \centering
        \includegraphics[width=0.48\linewidth]{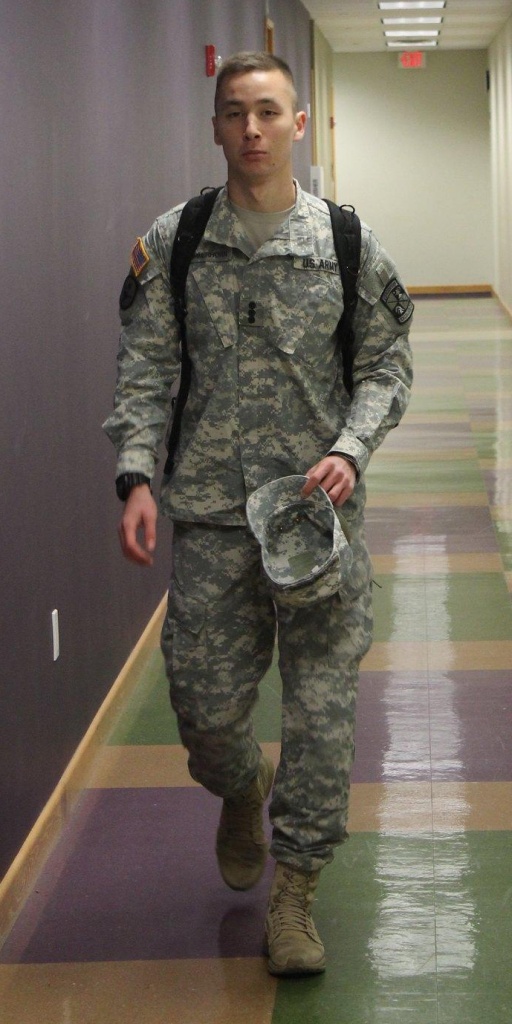}
        \includegraphics[width=0.48\linewidth]{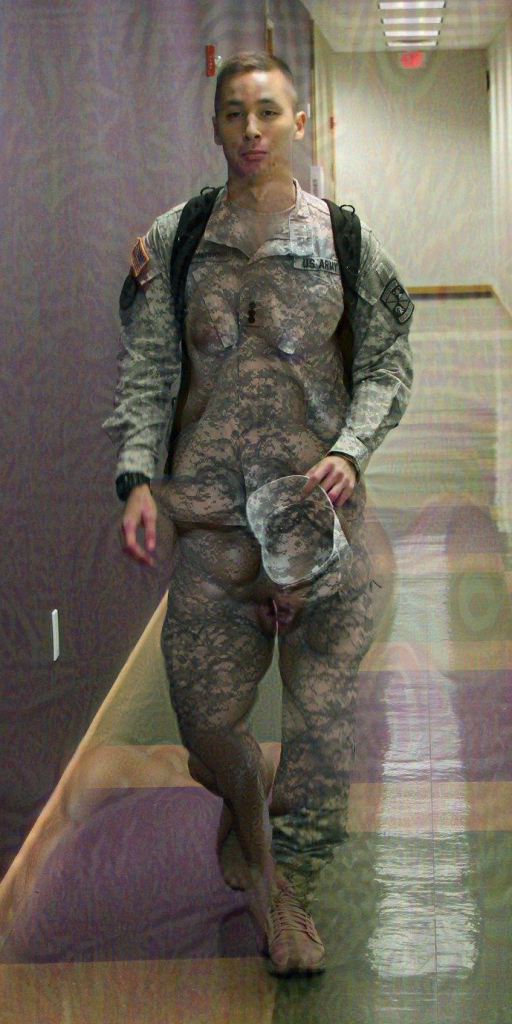}\\[-0.2em]
        \includegraphics[width=0.48\linewidth]{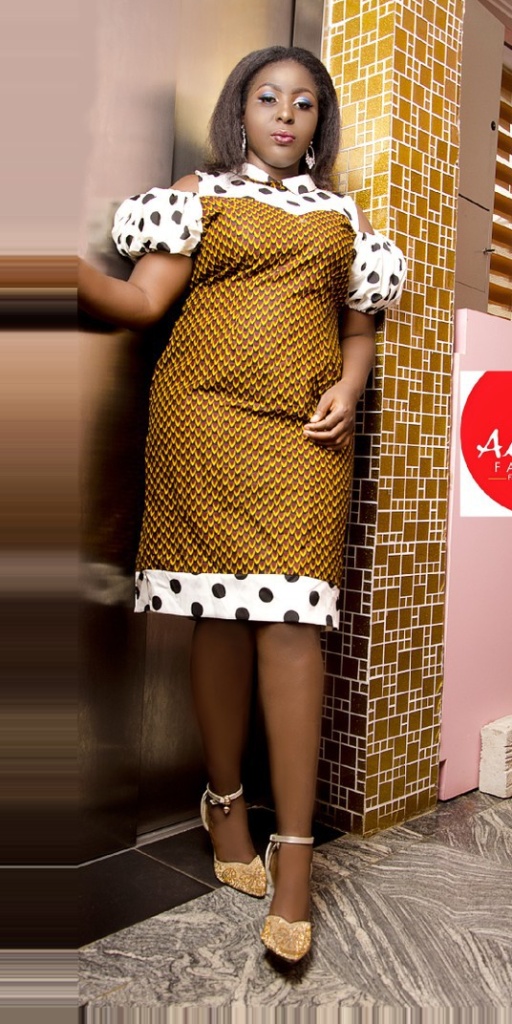}
        \includegraphics[width=0.48\linewidth]{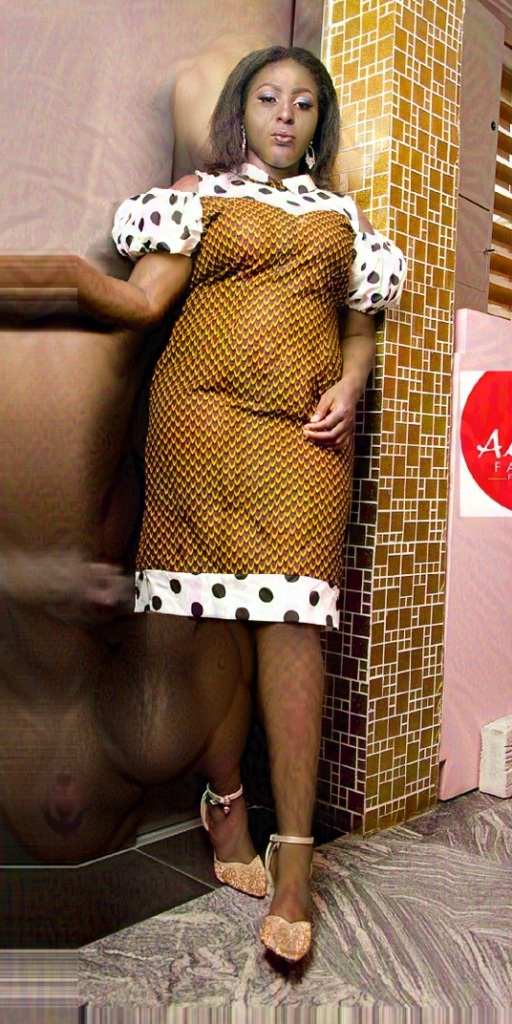}
        \caption{$||\delta||_\infty=8$}
    \end{subfigure}\hfill%
    \begin{subfigure}{0.48\linewidth}
        \centering
        \includegraphics[width=0.48\linewidth]{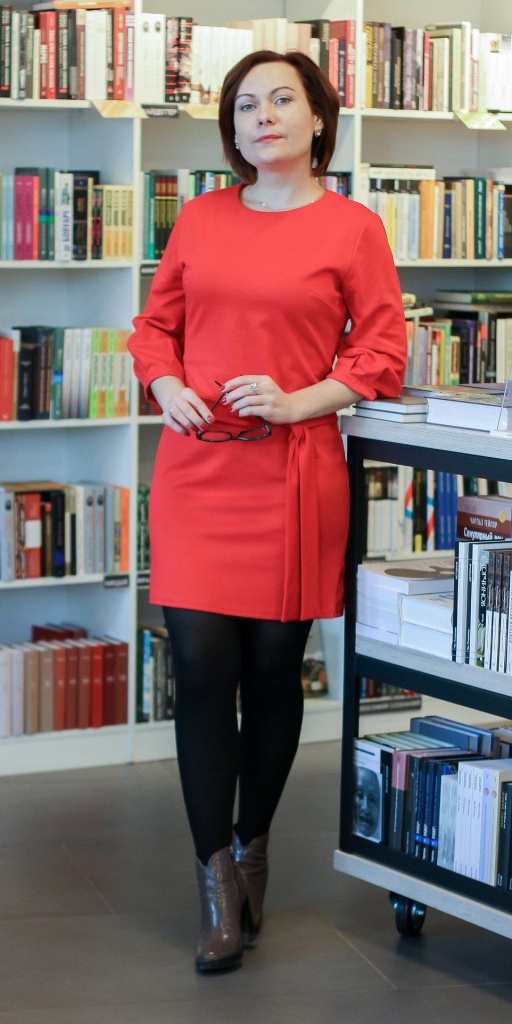}
        \includegraphics[width=0.48\linewidth]{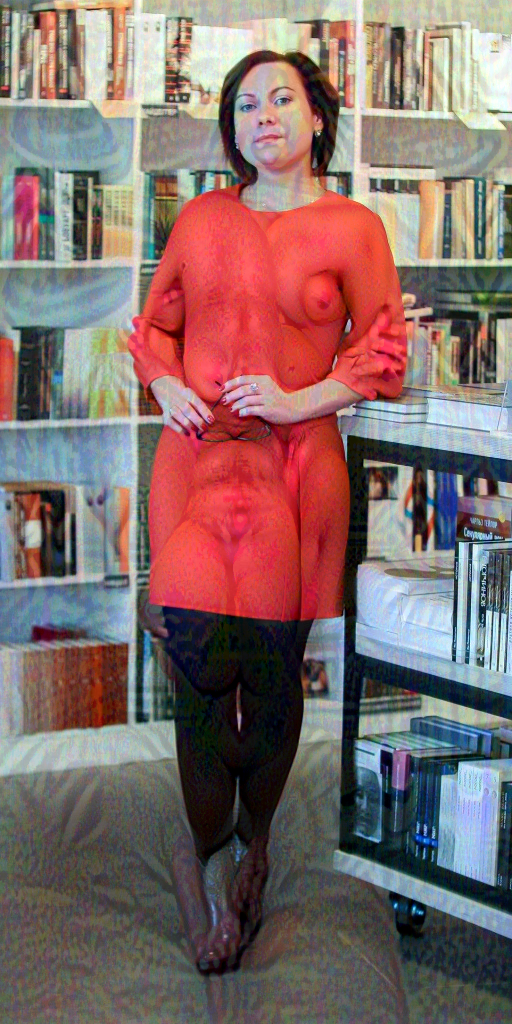}\\[-0.2em]
        \includegraphics[width=0.48\linewidth]{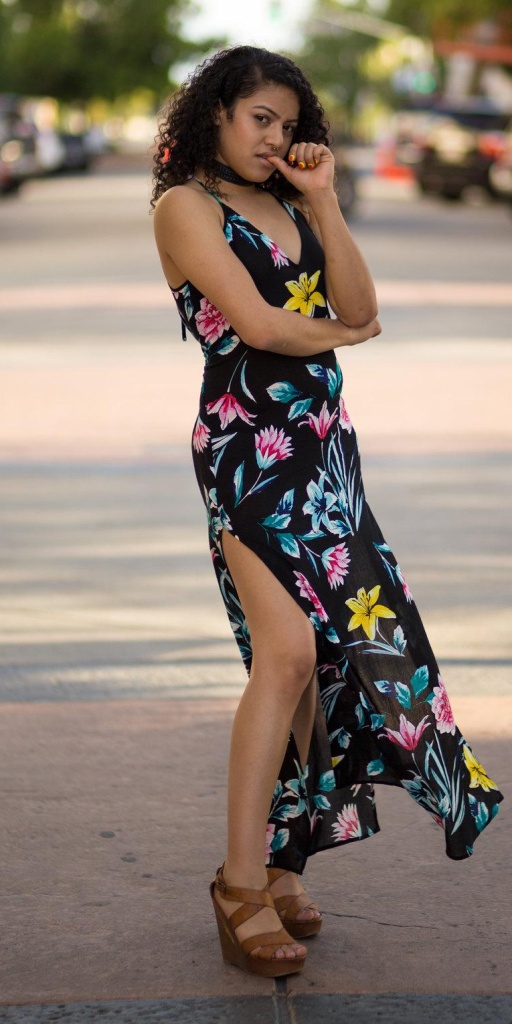}
        \includegraphics[width=0.48\linewidth]{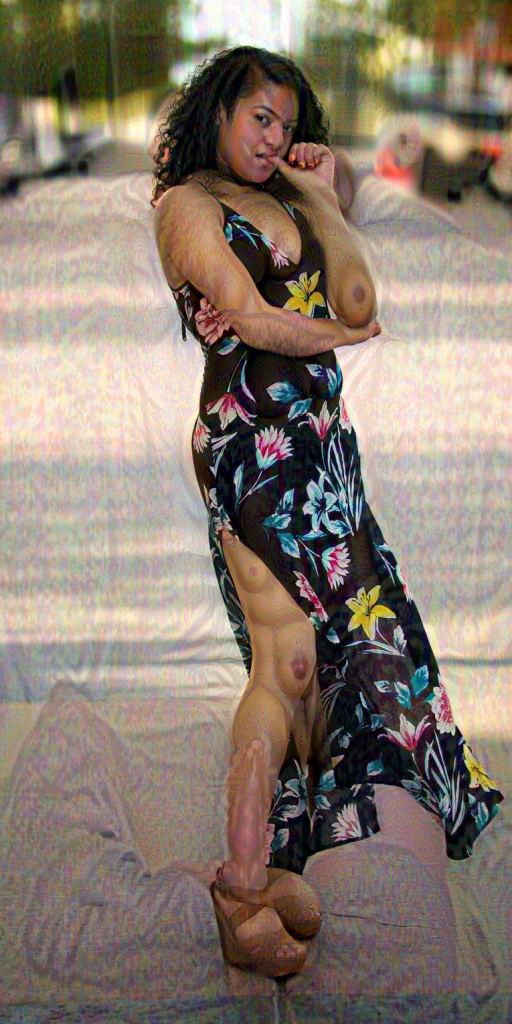}
        \caption{$||\delta||_\infty=16$}
    \end{subfigure}

    \caption{Additional qualitative examples of source images and their immunized counterparts.}
    \label{fig:qualitative-results-appendix}
\end{figure}

\Cref{fig:qualitative-results-appendix} shows additional examples of source images and their immunized counterparts at various perturbation budgets.

\subsection{Qualitative results on baselines}
\label{app:qualitative-baseline-results}
\begin{figure}[t]
    \centering
    \scriptsize
    \setlength{\tabcolsep}{1pt}
    \renewcommand{\arraystretch}{0.9}
    \begin{tabular*}{\columnwidth}{@{}>{\centering\arraybackslash}m{0.245\columnwidth}@{\extracolsep{\fill}}>{\centering\arraybackslash}m{0.245\columnwidth}>{\centering\arraybackslash}m{0.245\columnwidth}>{\centering\arraybackslash}m{0.245\columnwidth}@{}}
        \toprule
        Perturbed Image & Local Edit & API Edit (Clean Image) & API Edit (Perturbed Image) \\
        \midrule
        \includegraphics[width=0.24\columnwidth]{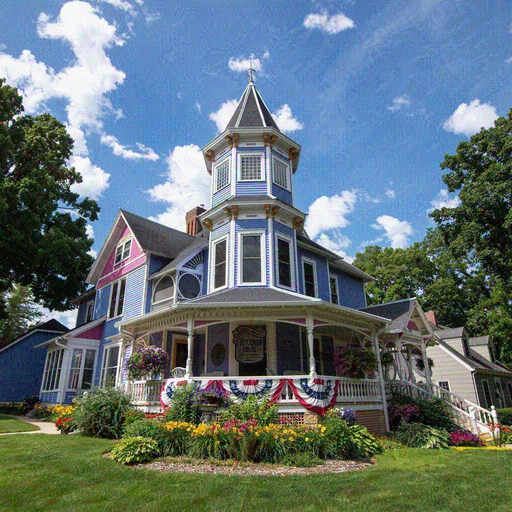} &
        \includegraphics[width=0.24\columnwidth]{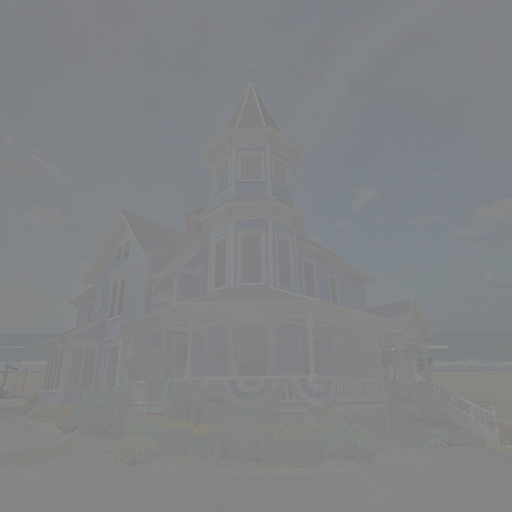} &
        \includegraphics[width=0.24\columnwidth]{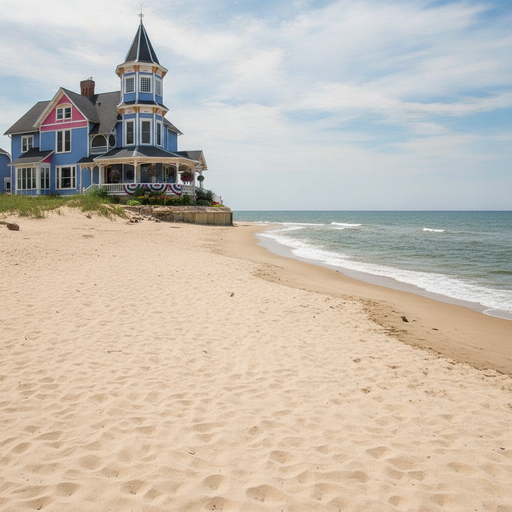} &
        \includegraphics[width=0.24\columnwidth]{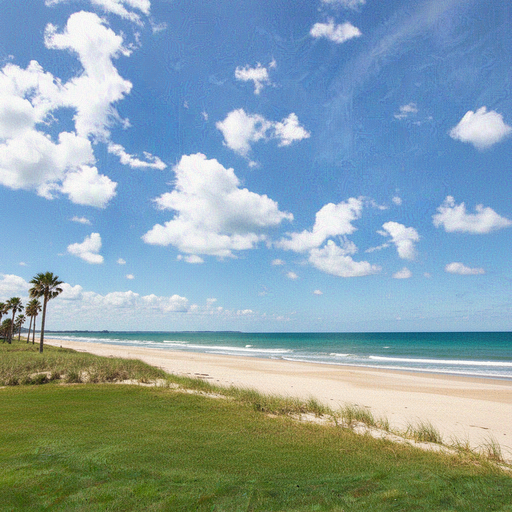} \\
        \multicolumn{4}{@{}>{\centering\arraybackslash}m{\columnwidth}@{}}{\tiny \textcolor{teal}{Successful immunization}, \textbf{Prompt:} \itshape Change the house to a coastal setting with a beach and ocean.} \\[0.35em]
        \includegraphics[width=0.24\columnwidth]{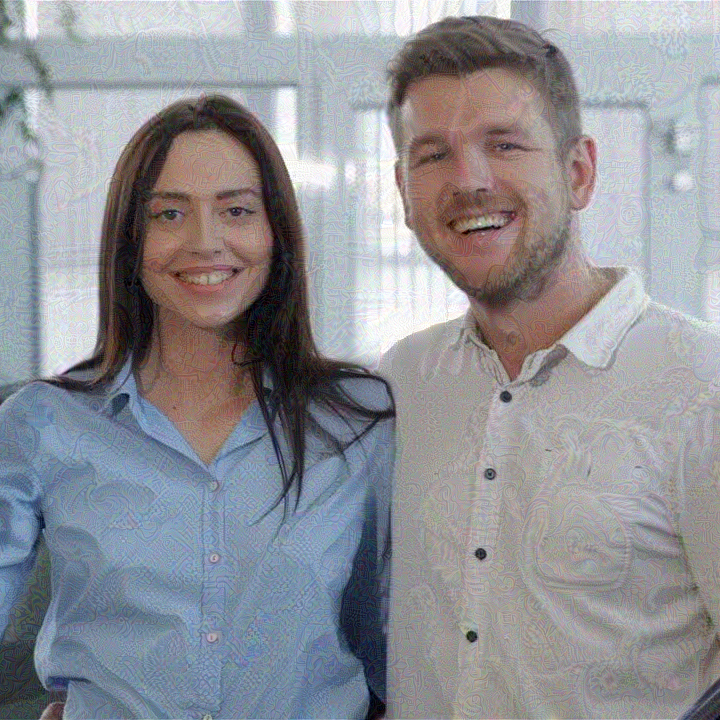} &
        \includegraphics[width=0.24\columnwidth]{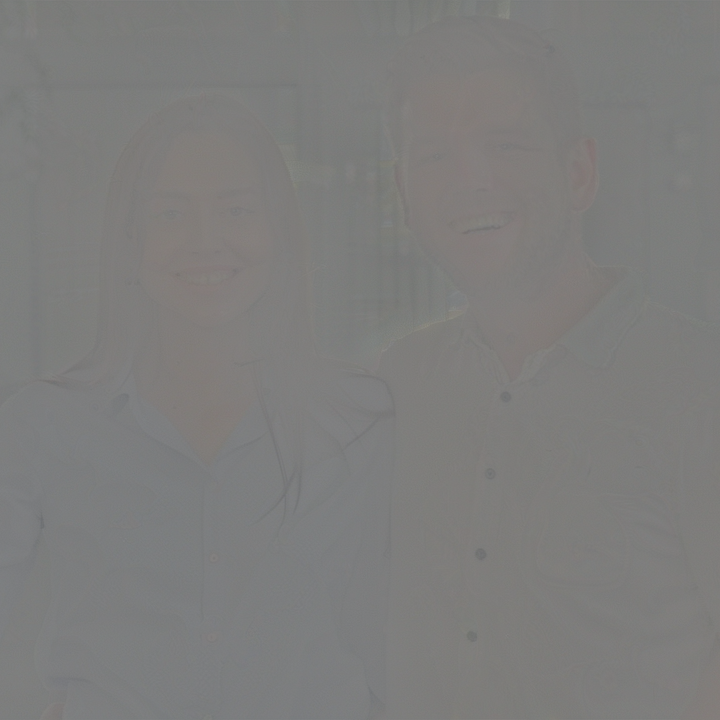} &
        \includegraphics[width=0.24\columnwidth]{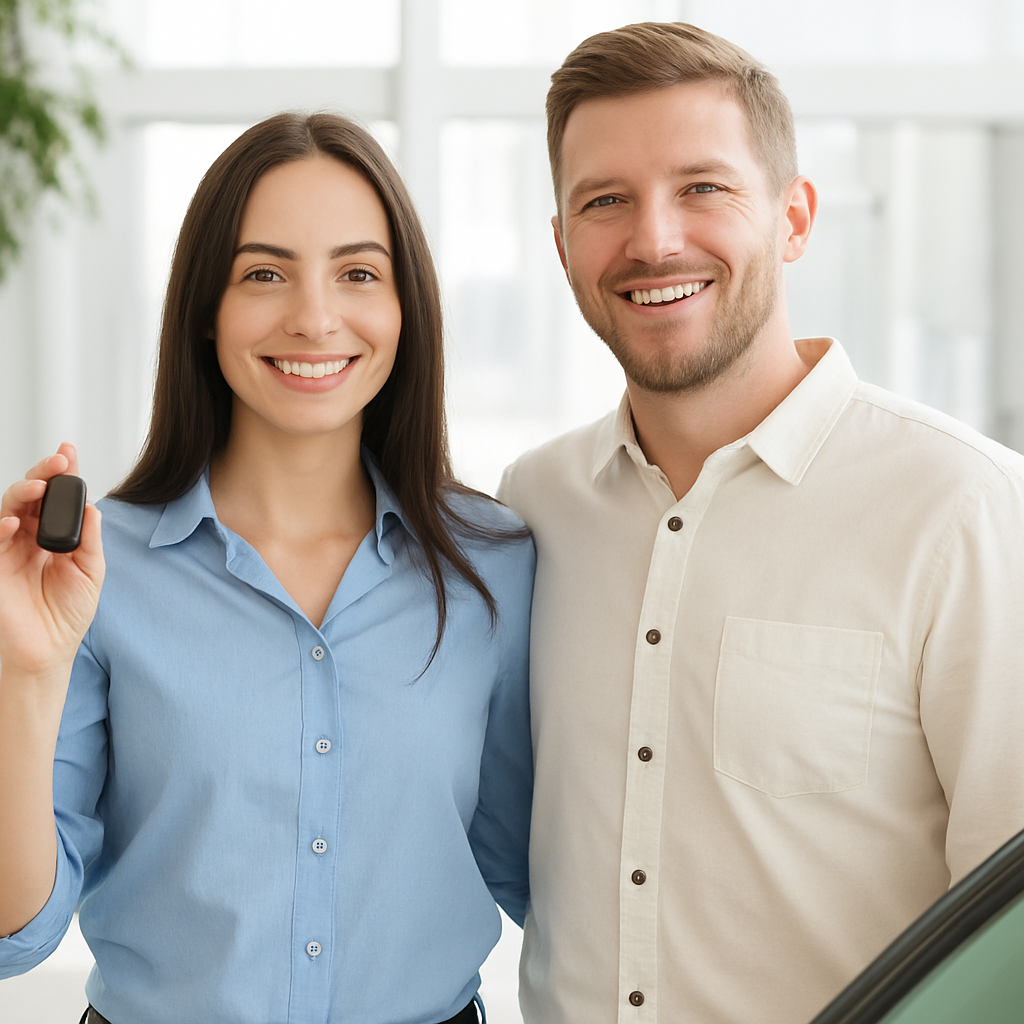} &
        \includegraphics[width=0.24\columnwidth]{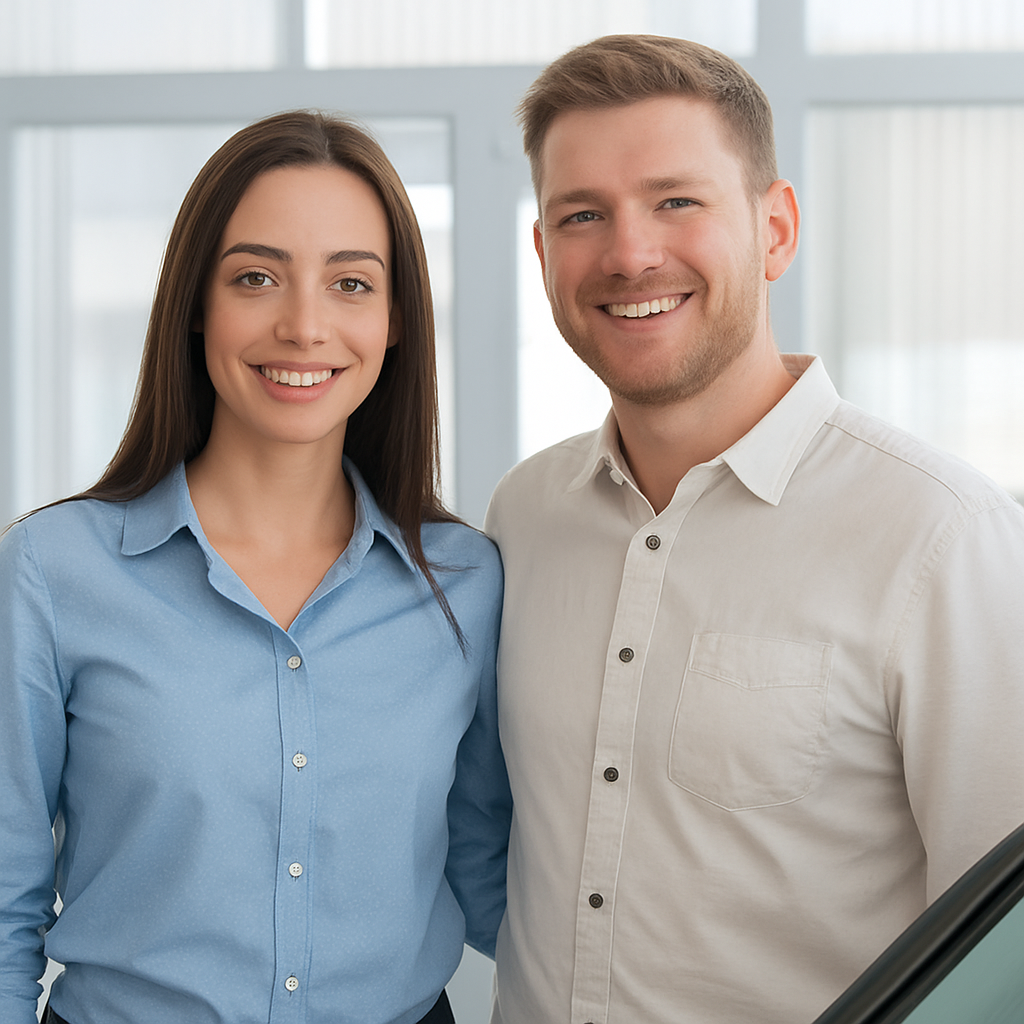} \\
        \multicolumn{4}{@{}>{\centering\arraybackslash}m{\columnwidth}@{}}{\tiny\textcolor{teal}{Successful immunization}, \textbf{Prompt:} \itshape Add a key to the left hand.} \\[0.35em]
        \includegraphics[width=0.24\columnwidth]{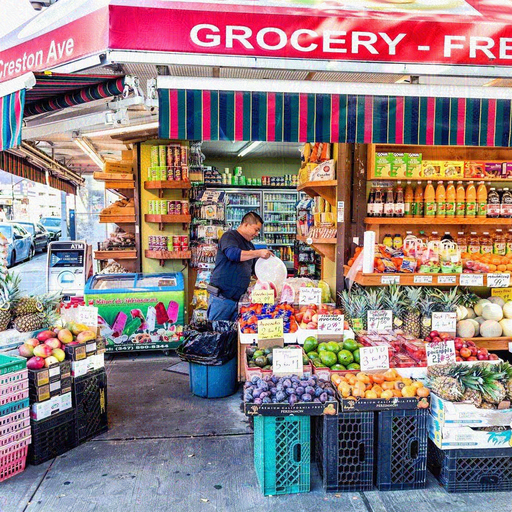} &
        \includegraphics[width=0.24\columnwidth]{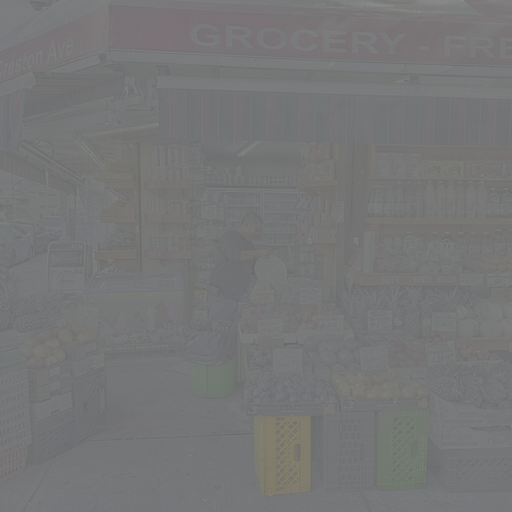} &
        \includegraphics[width=0.24\columnwidth]{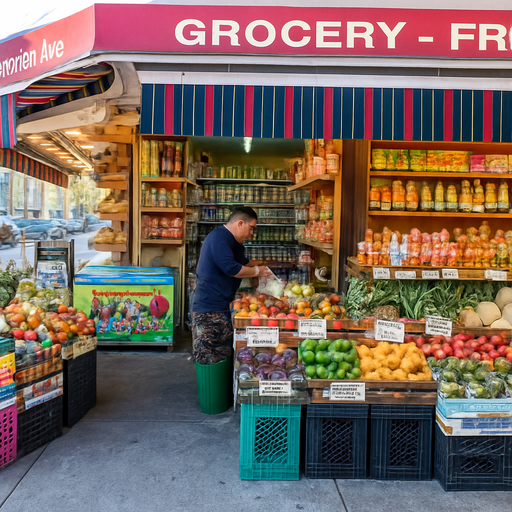} &
        \includegraphics[width=0.24\columnwidth]{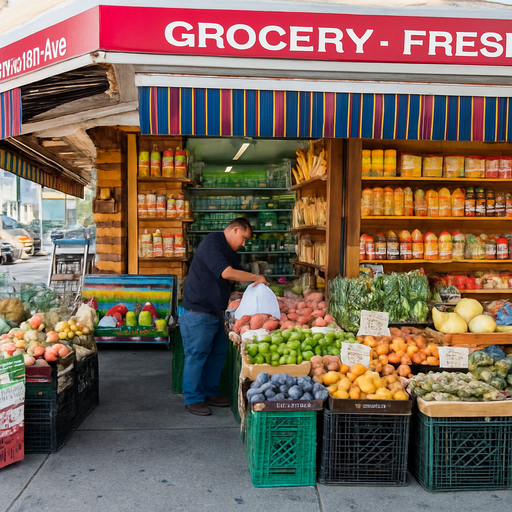} \\
        \multicolumn{4}{@{}>{\centering\arraybackslash}m{\columnwidth}@{}}{\tiny\textcolor{teal}{Successful immunization}, \textbf{Prompt:} \itshape Change the trash can next to the person to green.} \\[0.35em]
        \midrule
        \includegraphics[width=0.24\columnwidth]{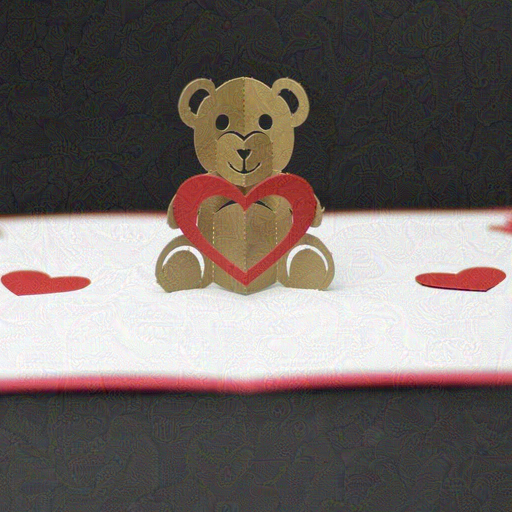} &
        \includegraphics[width=0.24\columnwidth]{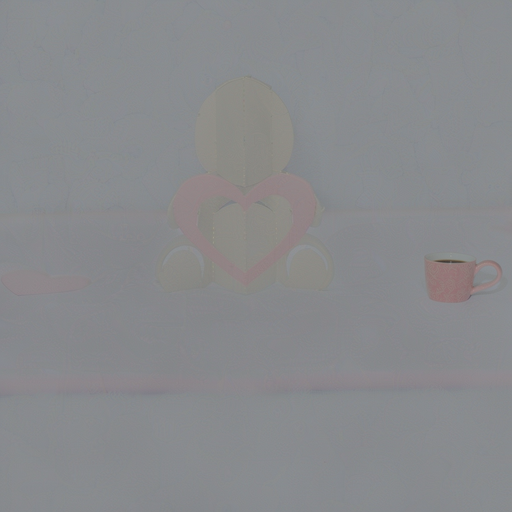} &
        \includegraphics[width=0.24\columnwidth]{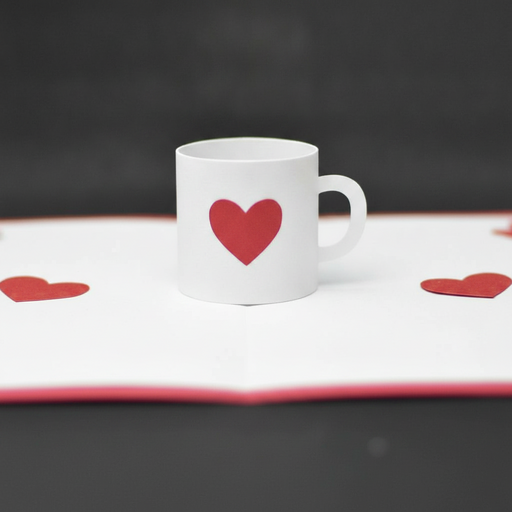} &
        \includegraphics[width=0.24\columnwidth]{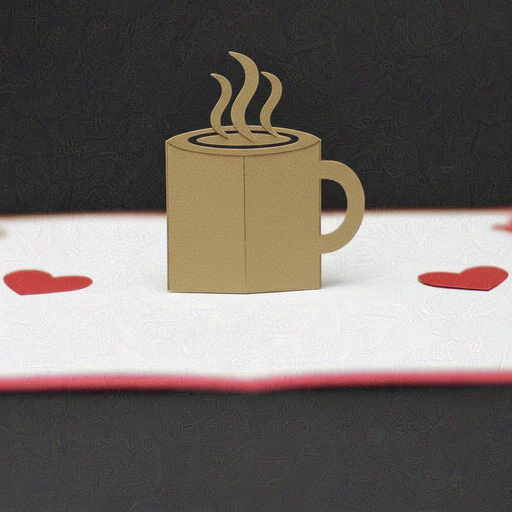} \\
        \multicolumn{4}{@{}>{\centering\arraybackslash}m{\columnwidth}@{}}{\tiny\textcolor{red}{Failed immunization}, \textbf{Prompt:} \itshape Replace the teddy bear holding a heart with a coffee mug.} \\[0.35em]
        \includegraphics[width=0.24\columnwidth]{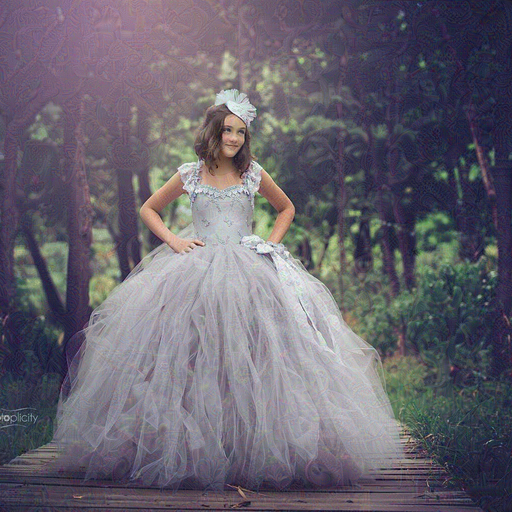} &
        \includegraphics[width=0.24\columnwidth]{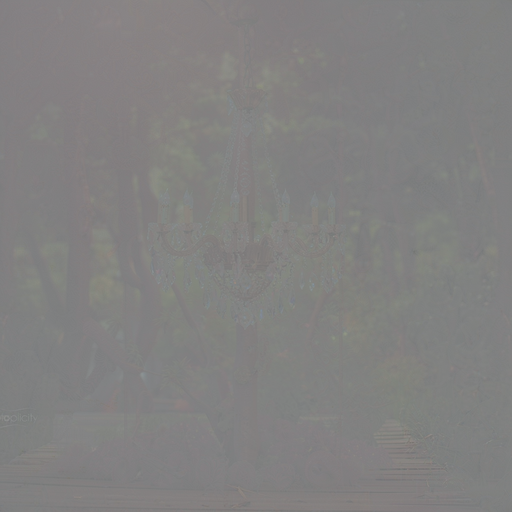} &
        \includegraphics[width=0.24\columnwidth]{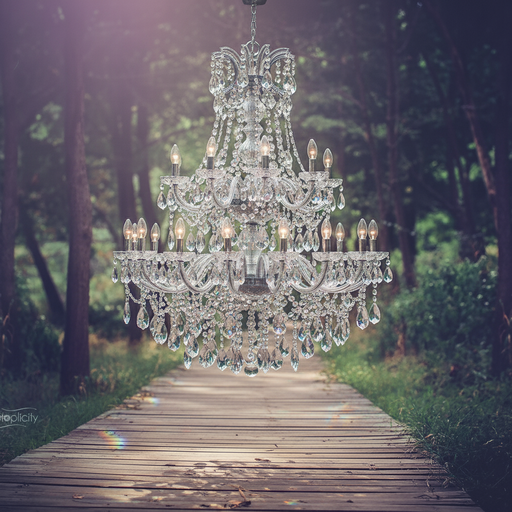} &
        \includegraphics[width=0.24\columnwidth]{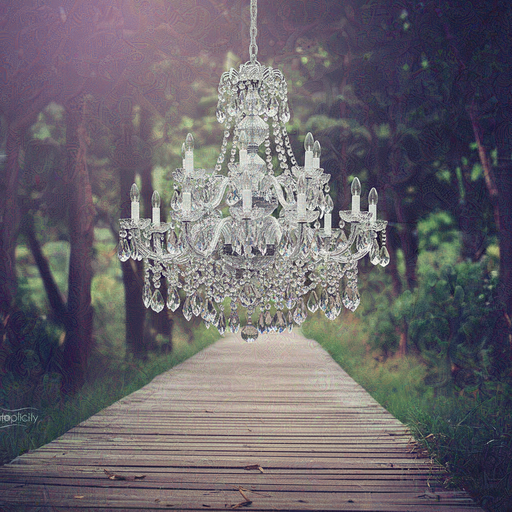} \\
        \multicolumn{4}{@{}>{\centering\arraybackslash}m{\columnwidth}@{}}{\tiny\textcolor{red}{Failed immunization}, \textbf{Prompt:} \itshape Replace the girl with a large crystal chandelier.} \\[0.35em]
        \includegraphics[width=0.24\columnwidth]{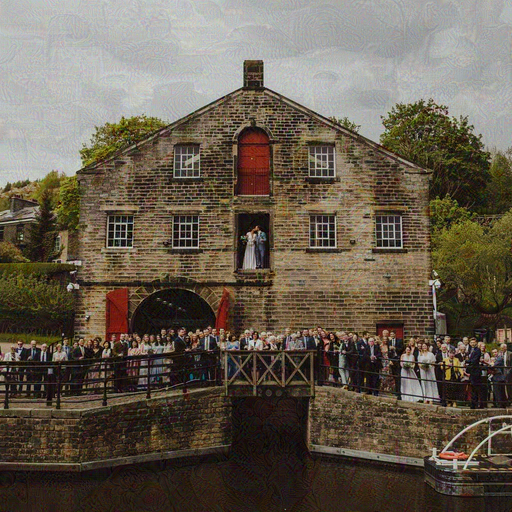} &
        \includegraphics[width=0.24\columnwidth]{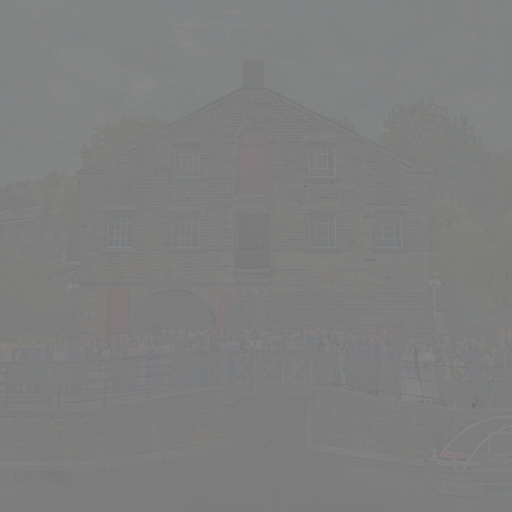} &
        \includegraphics[width=0.24\columnwidth]{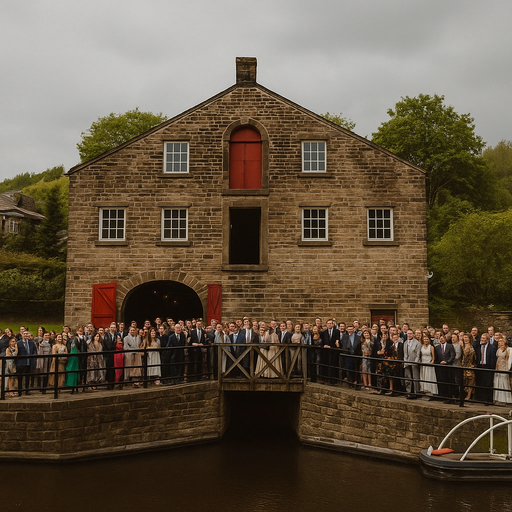} &
        \includegraphics[width=0.24\columnwidth]{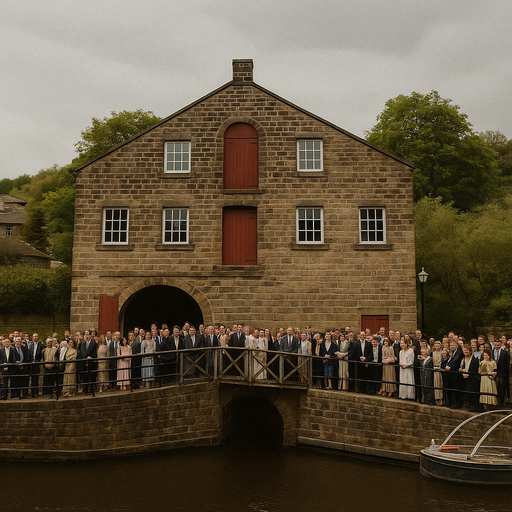} \\
        \multicolumn{4}{@{}>{\centering\arraybackslash}m{\columnwidth}@{}}{\tiny\textcolor{red}{Failed immunization}, \textbf{Prompt:} \itshape Remove the couple on the second floor of the house.} \\[0.35em]
        \bottomrule
    \end{tabular*}
    \caption{\textbf{Qualitative PhotoGuard~\cite{salman2023photoguard} examples}. Each row shows the perturbed image, the local white-box edit, the API edit of the clean image, and the API edit of the perturbed image.  The top three rows are successfully immunized, where the VLM judge determines that the perturbed API edit does not follow the instruction, while the bottom three rows are immunization failures. We see that the VLM-as-a-judge  is consistent with human judgment for the immunized results on the rightmost column.}
    \label{fig:baseline-photoguard-qualitative-results}
\end{figure}

\begin{figure}[t]
    \centering
    \scriptsize
    \setlength{\tabcolsep}{1pt}
    \renewcommand{\arraystretch}{0.9}
    \begin{tabular*}{\columnwidth}{@{}>{\centering\arraybackslash}m{0.245\columnwidth}@{\extracolsep{\fill}}>{\centering\arraybackslash}m{0.245\columnwidth}>{\centering\arraybackslash}m{0.245\columnwidth}>{\centering\arraybackslash}m{0.245\columnwidth}@{}}
        \toprule
        Perturbed Image & Local Edit & API Edit (Clean Image) & API Edit (Perturbed Image) \\
        \midrule
        \includegraphics[width=0.24\columnwidth]{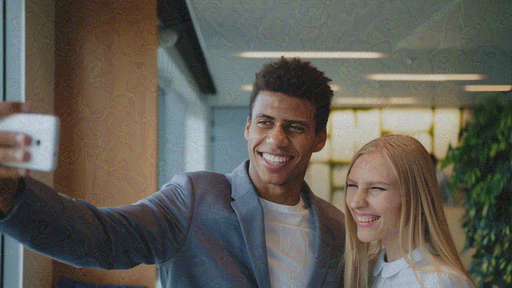} &
        \includegraphics[width=0.24\columnwidth]{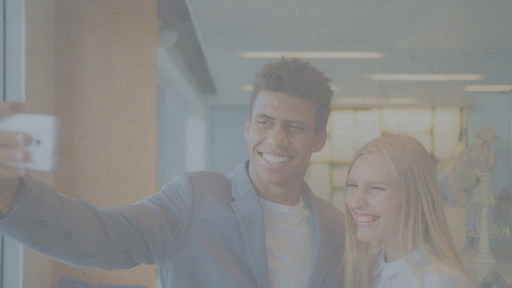} &
        \includegraphics[width=0.24\columnwidth]{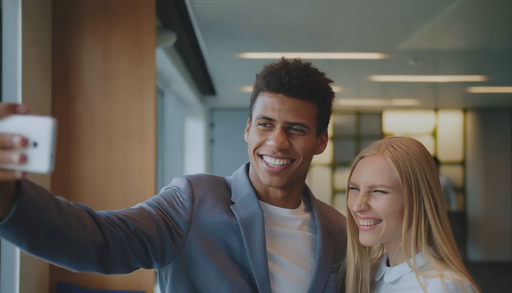} &
        \includegraphics[width=0.24\columnwidth]{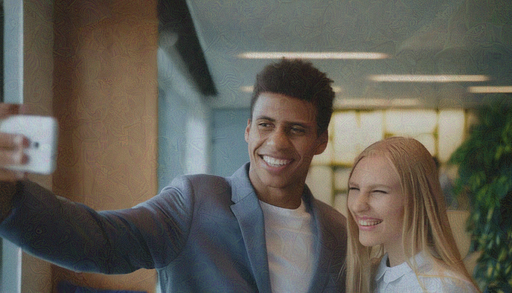} \\
        \multicolumn{4}{@{}>{\centering\arraybackslash}m{\columnwidth}@{}}{\tiny\textcolor{teal}{Successful immunization}, \textbf{Prompt:} \itshape Remove the plant and adjust the suit to a darker shade of blue.} \\[0.35em]
        \includegraphics[width=0.24\columnwidth]{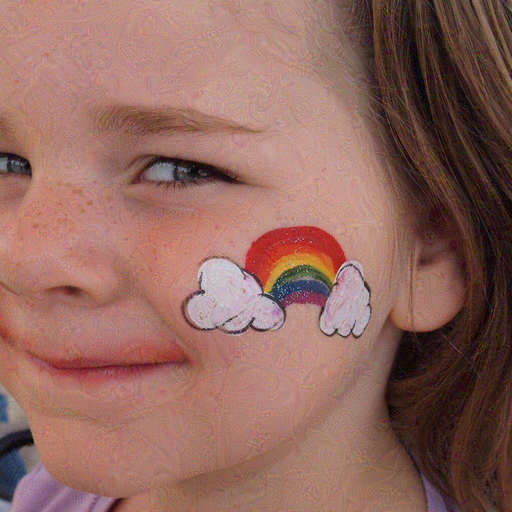} &
        \includegraphics[width=0.24\columnwidth]{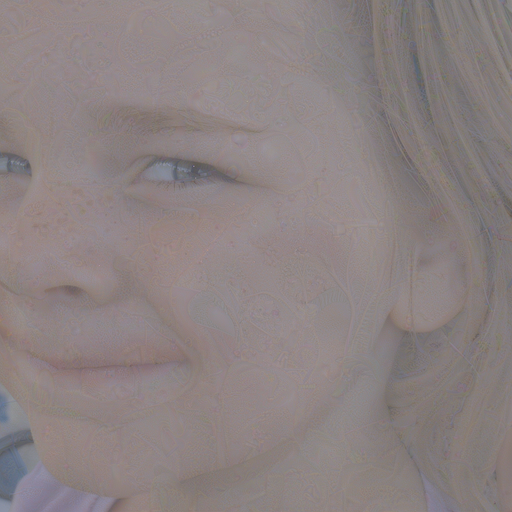} &
        \includegraphics[width=0.24\columnwidth]{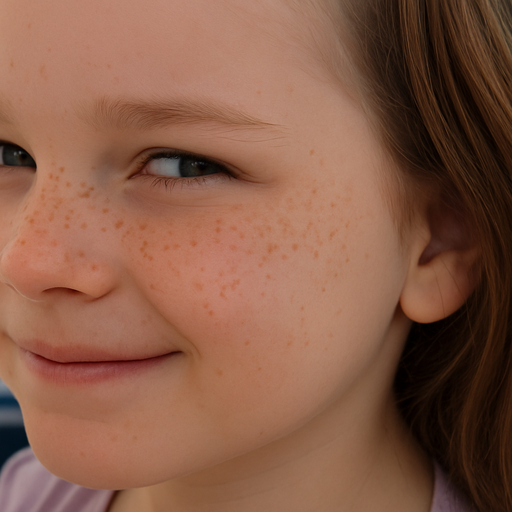} &
        \includegraphics[width=0.24\columnwidth]{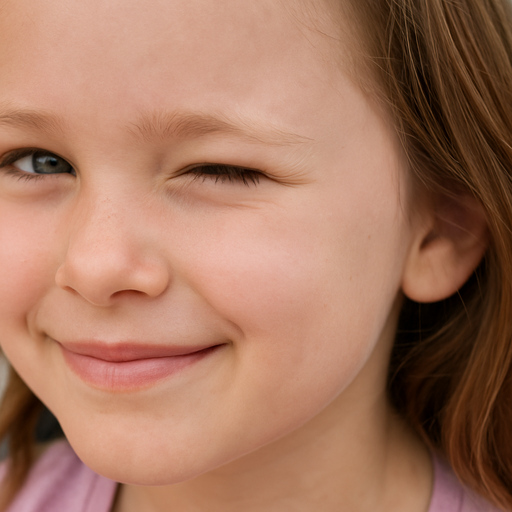} \\
        \multicolumn{4}{@{}>{\centering\arraybackslash}m{\columnwidth}@{}}{\tiny\textcolor{teal}{Successful immunization}, \textbf{Prompt:} \itshape Remove the rainbow and cloud face paint.} \\[0.35em]
        \includegraphics[width=0.24\columnwidth]{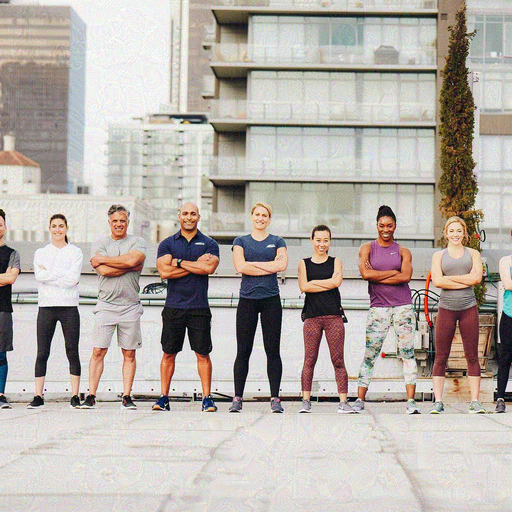} &
        \includegraphics[width=0.24\columnwidth]{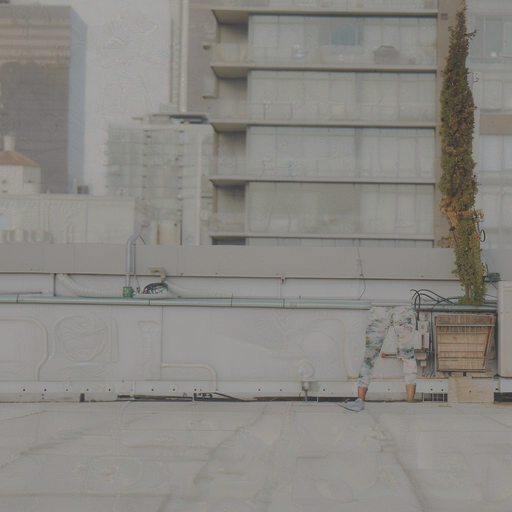} &
        \includegraphics[width=0.24\columnwidth]{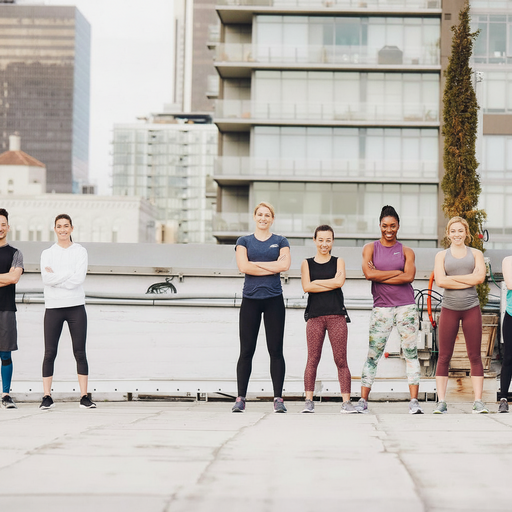} &
        \includegraphics[width=0.24\columnwidth]{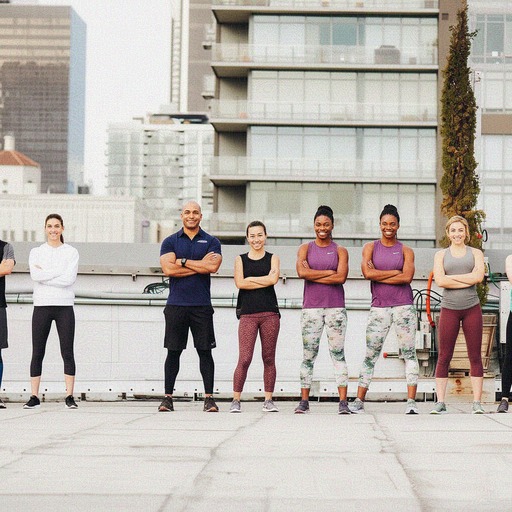} \\
        \multicolumn{4}{@{}>{\centering\arraybackslash}m{\columnwidth}@{}}{\tiny\textcolor{teal}{Successful immunization}, \textbf{Prompt:} \itshape Remove all people wearing shorts.} \\[0.35em]
        \midrule
        \includegraphics[width=0.24\columnwidth]{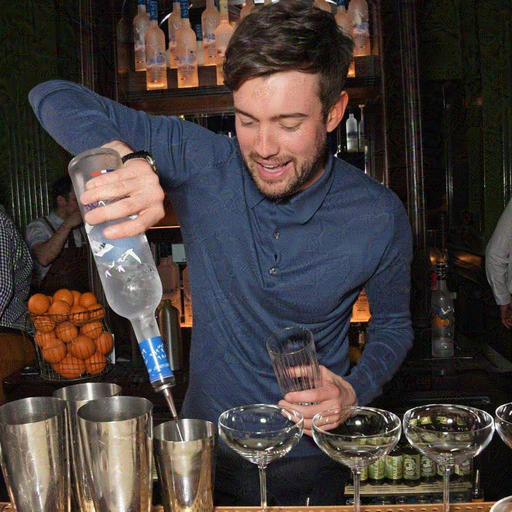} &
        \includegraphics[width=0.24\columnwidth]{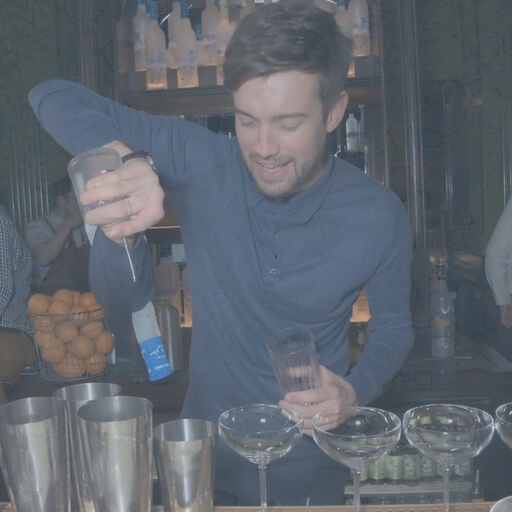} &
        \includegraphics[width=0.24\columnwidth]{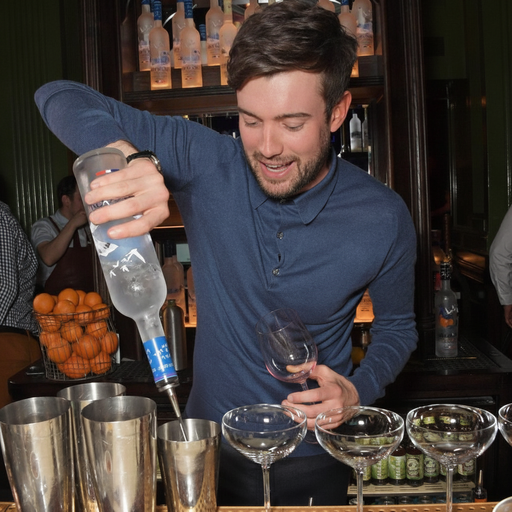} &
        \includegraphics[width=0.24\columnwidth]{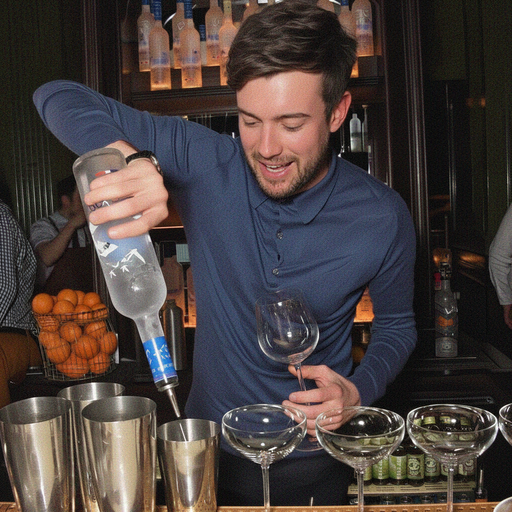} \\
        \multicolumn{4}{@{}>{\centering\arraybackslash}m{\columnwidth}@{}}{\tiny\textcolor{red}{Failed immunization}, \textbf{Prompt:} \itshape Change the cup in the man's hand into a wine glass.} \\[0.35em]
        \includegraphics[width=0.24\columnwidth]{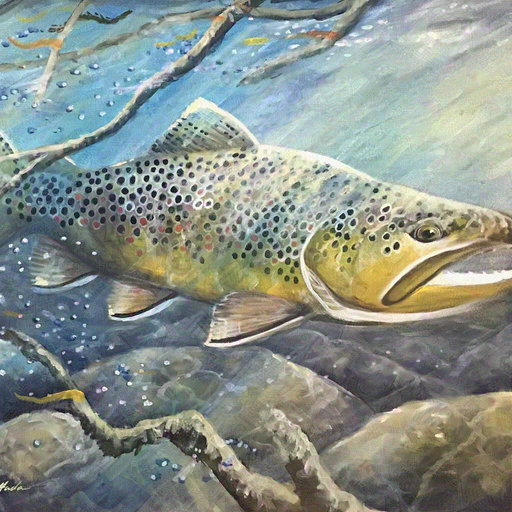} &
        \includegraphics[width=0.24\columnwidth]{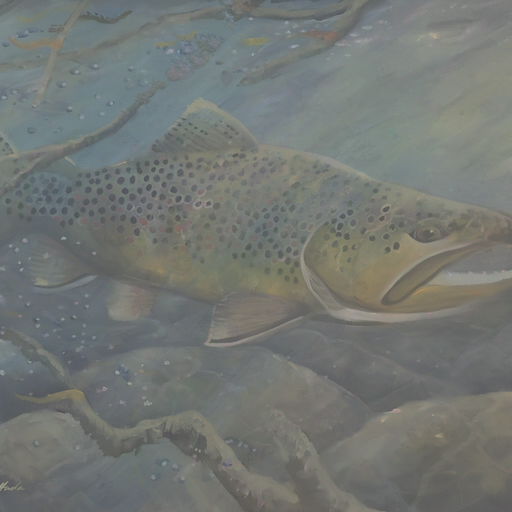} &
        \includegraphics[width=0.24\columnwidth]{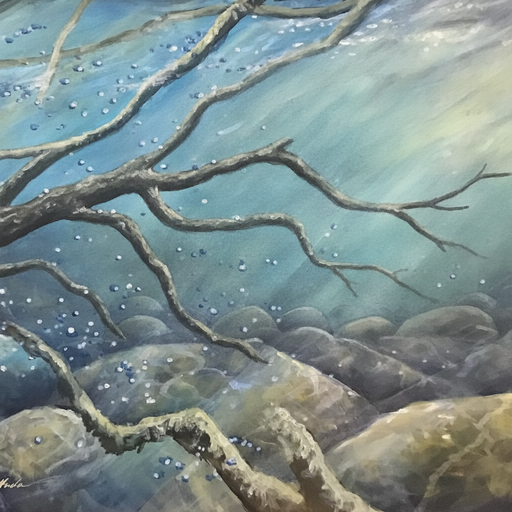} &
        \includegraphics[width=0.24\columnwidth]{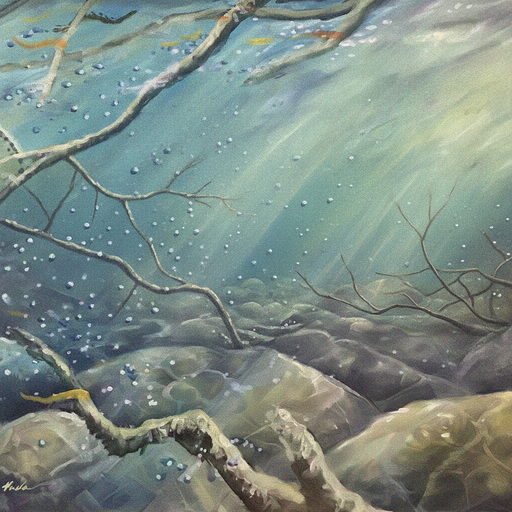} \\
        \multicolumn{4}{@{}>{\centering\arraybackslash}m{\columnwidth}@{}}{\tiny\textcolor{red}{Failed immunization}, \textbf{Prompt:} \itshape Remove the brown trout fish in the center.} \\[0.35em]
        \includegraphics[width=0.24\columnwidth]{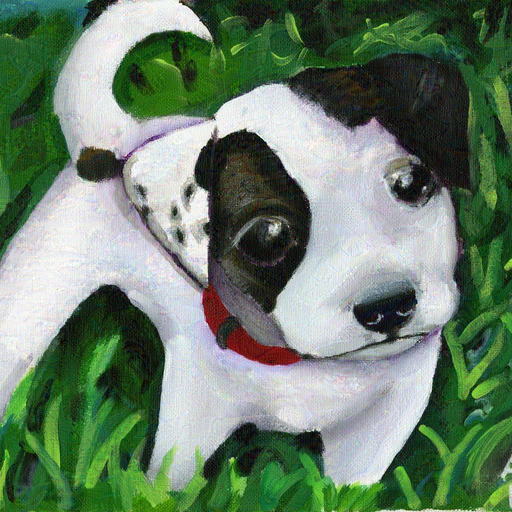} &
        \includegraphics[width=0.24\columnwidth]{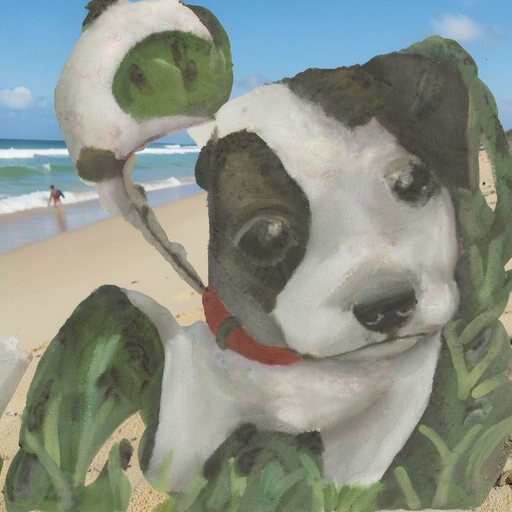} &
        \includegraphics[width=0.24\columnwidth]{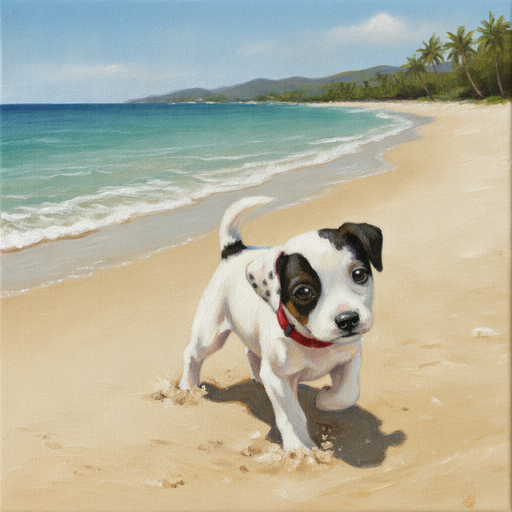} &
        \includegraphics[width=0.24\columnwidth]{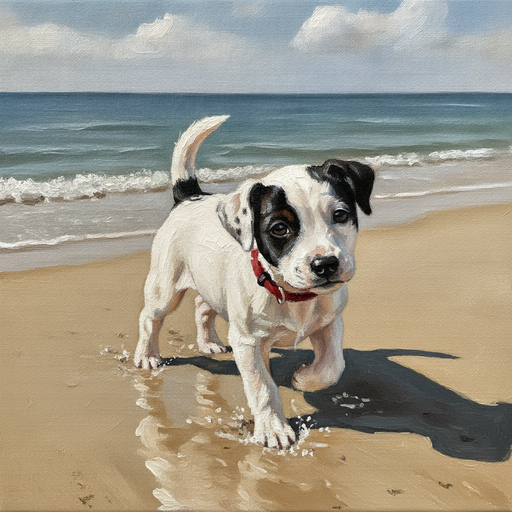} \\
        \multicolumn{4}{@{}>{\centering\arraybackslash}m{\columnwidth}@{}}{\tiny\textcolor{red}{Failed immunization}, \textbf{Prompt:} \itshape Change the grass background to a sandy beach with the puppy near water.} \\[0.35em]
        \bottomrule
\end{tabular*}
    \caption{\textbf{Qualitative TDAE~\cite{zhang2026towards} examples} Each row shows the perturbed image, the local white-box edit, the API edit of the clean image, and the API edit of the perturbed image.  The top three rows are successfully immunized, where the VLM judge determines that the perturbed API edit does not follow the instruction, while the bottom three rows are immunization failures. We see that the VLM-as-a-judge  is consistent with human judgment for the immunized results on the rightmost column.}
    \label{fig:baseline-tdae-qualitative-results}
\end{figure}

From our main results, we see that even baselines that explicitly target transferable immunization, such as TDAE~\cite{zhang2026towards} do not protect against image editing by proprietary models. One could contest that the low immunization rates of the baselines are a consequence of our VLM-as-a-judge setup. To address this, we also show qualitative examples of successfully and unsuccessfully immunized images (as determined by the VLM judge) in Figs.~\ref{fig:baseline-photoguard-qualitative-results} and \ref{fig:baseline-tdae-qualitative-results}. In each of the figures, we present three successfully immunized images (top three rows) and three immunization failures.  As we see, images are successfully immunized against the local Flux model, leading to degraded output quality. However, on the API models, the output is high quality, albeit not following the instruction in certain cases.

\subsection{Compound Immunization with PhotoGuard}
\label{app:mirage-with-pg}

\begin{figure*}[t]
    \centering
    \scriptsize
    \setlength{\tabcolsep}{1pt}
    \renewcommand{\arraystretch}{0.95}
    \begin{tabular*}{\textwidth}{@{}>{\centering\arraybackslash}m{0.15\textwidth}@{\extracolsep{\fill}}|>{\centering\arraybackslash}m{0.12\textwidth}|>{\centering\arraybackslash}m{0.12\textwidth}>{\centering\arraybackslash}m{0.12\textwidth}>{\centering\arraybackslash}m{0.12\textwidth}|>{\centering\arraybackslash}m{0.12\textwidth}>{\centering\arraybackslash}m{0.12\textwidth}>{\centering\arraybackslash}m{0.07\textwidth}@{}}
        \toprule
        Prompt & Original & \multicolumn{3}{c}{PhotoGuard} & \multicolumn{3}{c}{+\methodname{}} \\
        & & Perturbed & Locally edited & API edited & Perturbed & Locally edited & API edited \\
        \midrule
        Edit some mountains in the background. &
        \includegraphics[width=0.12\textwidth]{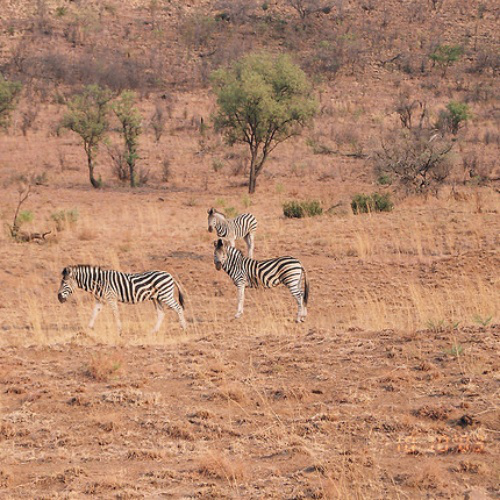} &
        \includegraphics[width=0.12\textwidth]{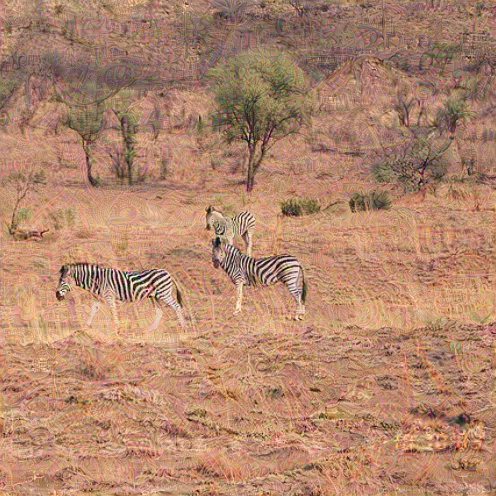} &
        \includegraphics[width=0.12\textwidth]{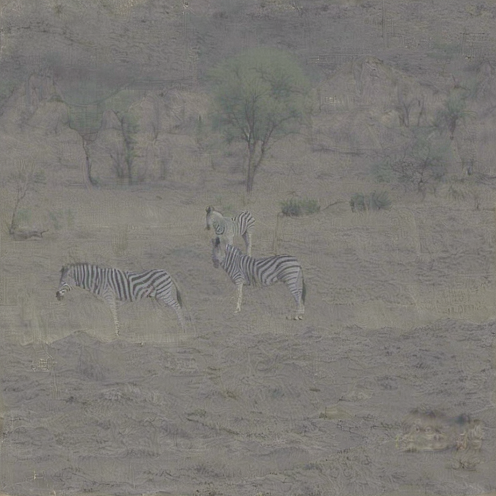} &
        \includegraphics[width=0.12\textwidth]{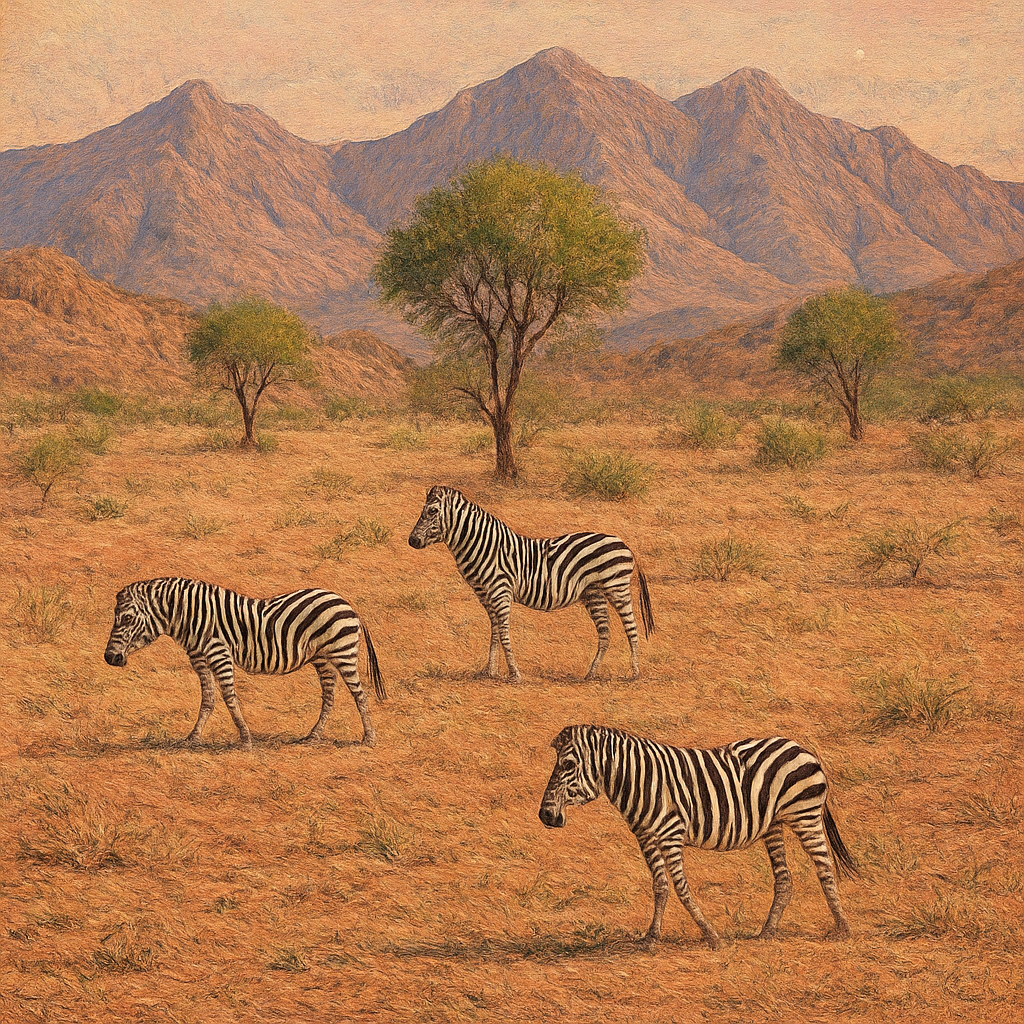} &
        \includegraphics[width=0.12\textwidth]{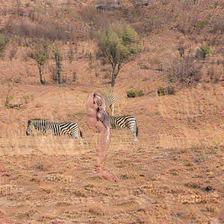} &
        \includegraphics[width=0.12\textwidth]{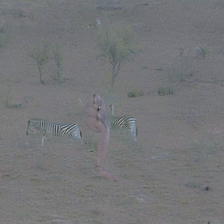} &
        Refused \\
        Replace the baseball bat with a laser sword. &
        \includegraphics[width=0.12\textwidth]{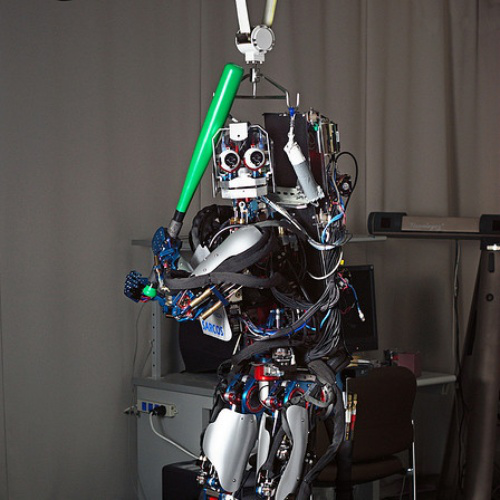} &
        \includegraphics[width=0.12\textwidth]{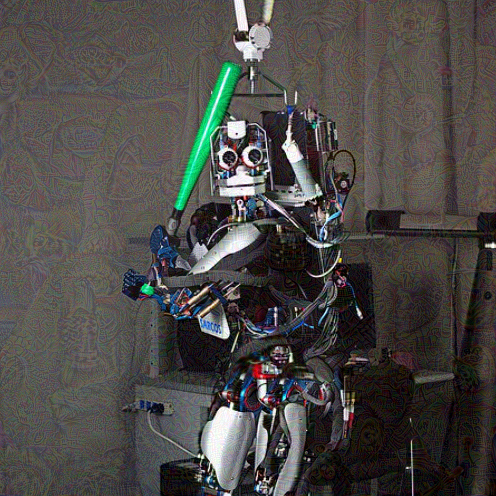} &
        \includegraphics[width=0.12\textwidth]{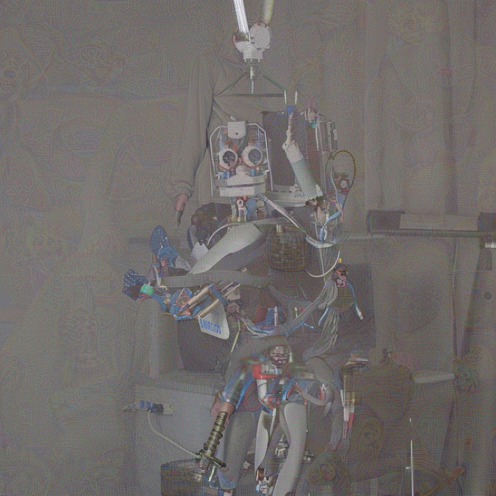} &
        \includegraphics[width=0.12\textwidth]{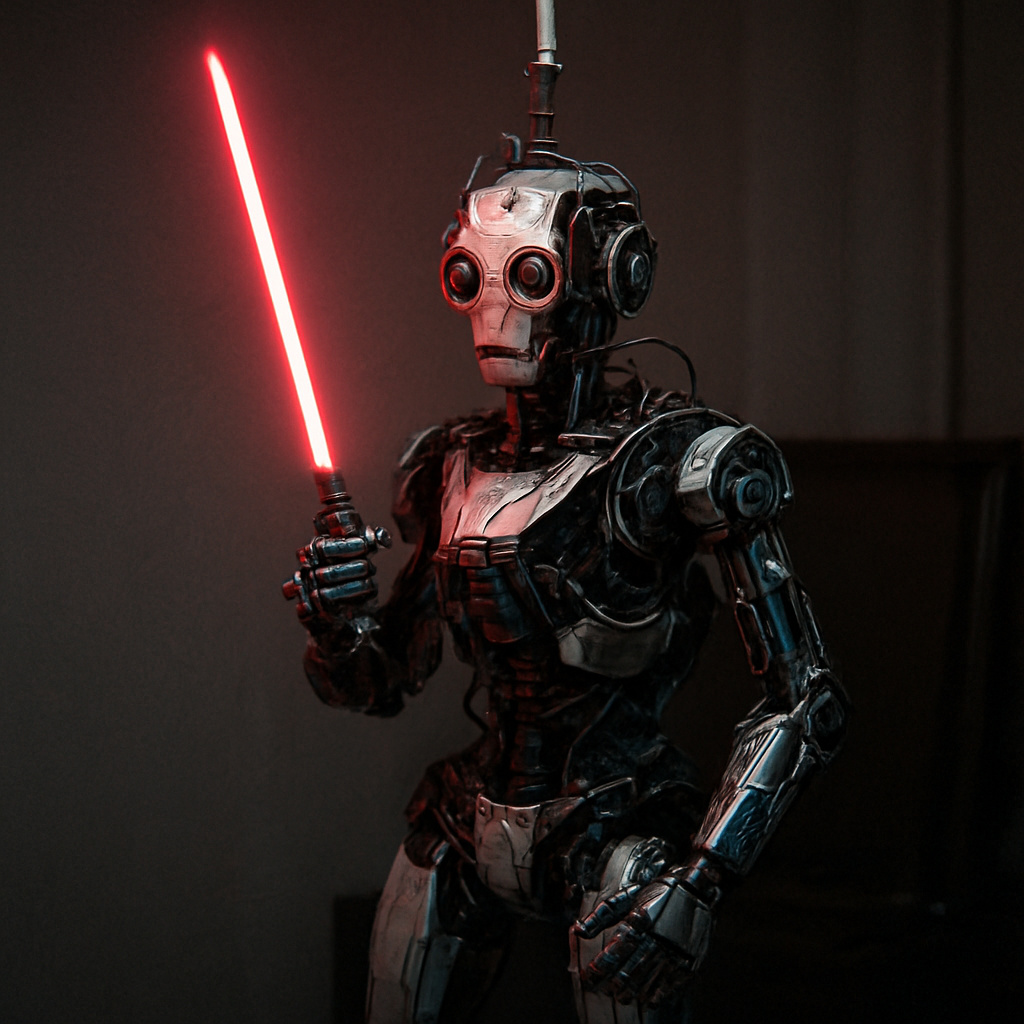} &
        \includegraphics[width=0.12\textwidth]{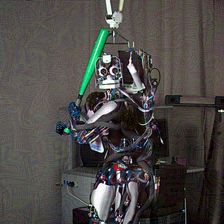} &
        \includegraphics[width=0.12\textwidth]{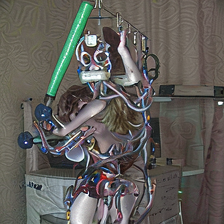} &
        Refused \\
        \bottomrule
    \end{tabular*}
    \caption{Qualitative comparison between PhotoGuard (PG) and \methodname{} on the same edit prompts demonstrates that combining PG and \methodname{} in the perturbation objective provides protection against both open-weight models and black-box APIs, simultaneously and with a single perturbation.}
    \label{fig:mirage-photoguard-qualitative-results}
\end{figure*}

As we saw in \cref{sec:results-attack}, a strong adversary with access to a local diffusion model could easily bypass \methodname and edit the image locally. If the immunizer knew what local model might be used to edit the images, they could add a PhotoGuard-style objective with the desired local model as one of the surrogates during the immunization procedure. We implement this, and present qualitative results in \cref{fig:mirage-photoguard-qualitative-results}. In each row, we show the original image the PhotoGuard perturbation, local and API edits on the PhotoGuard protected image. PhotoGuard can protect against editing by the local model, but not against API edits. We then show perturbed image found by \methodname + PhotoGuard, where a Flux-4B model is used as one of the surrogate models in optimizing \cref{eqn:main-opt}. We find that the perturbed image is refused by the image editing APIs. It also leads to degraded image edits by the local model, getting the best of both worlds.

\end{document}